\documentclass[aps,preprint]{revtex4}

\usepackage{amsfonts}
\usepackage{amsmath}
\usepackage{amssymb}
\usepackage{subfigure}
\usepackage{graphicx}
\usepackage{epstopdf}
\usepackage{epsfig}
\usepackage{float}
\usepackage{color}
\usepackage{ulem}
\usepackage{cancel}
\usepackage{mathrsfs}
\usepackage[colorlinks,linkcolor=blue,citecolor=blue,urlcolor=blue]{hyperref}
\usepackage{url}
\usepackage{cleveref}
\usepackage{ulem}

\begin{document}

\title{Thermodynamics and phase transition of BTZ black hole in a cavity}
\author{Yuchen Huang}
\email{huangyuchen@stu.scu.edu.cn}
\author{Jun Tao}
\email{taojun@scu.edu.cn}
	
\affiliation{Center for Theoretical Physics, College of Physics, Sichuan University, Chengdu, 610065, China}

\begin{abstract}

In this paper, we study the thermodynamics and phase transition of a BTZ black hole in a finite space region, namely a cavity. By imposing a temperature-fixed boundary condition on the wall of the cavity and evaluating the Euclidean action, we derive the thermodynamic quantities and then construct the first law of thermodynamics for a static and neutral BTZ black hole, a rotating BTZ black hole and a charged BTZ black hole, respectively. We prove that heat capacities of these three types of black holes are always non-negative. Considering a grand canonical ensemble, we find that the non-extreme rotating black hole and the charged black hole are locally thermodynamically stable by calculating the Hessian matrix of their internal energy. At the phase transition level, it shows that for the static and neutral BTZ black hole, the phase transition only exists between thermal AdS$_3$ spacetime and the black hole. The temperature where the phase transition occurs is only determined by the cavity radius. For rotating and charged cases, there may exist an extra second-order phase transition between the black hole and the black hole-cavity merger state. The phase structure of a BTZ black hole in a cavity shows strong dissimilarities from that without the cavity.

\end{abstract}

\maketitle

\section{Introduction}

The heat capacity of a Schwarzschild black hole in asymptotically flat spacetime is negative, which leads the system to become thermodynamically unstable. It was then proposed by Hawking and Page that the Einstein equation with a negative cosmological constant admits a black hole solution \cite{Hawking:1982dh}, where the Anti-de Sitter (AdS) space acts as a box of finite volume to make the canonical ensemble well defined and the Hawking-Page phase transition (the phase transition between thermal AdS space and the black hole) was found. Another method to thermally stabilize a black hole is to place it in a cavity \cite{York:1986it,Brown:1994gs}. Under this new boundary condition, York found that there exists a Hawking-Page-like phase transition. Afterwards, the charged black hole was investigated in a cavity in a grand canonical ensemble \cite{Braden:1990hw} and a canonical ensemble \cite{Lundgren:2006kt}, where the similar phase transitions were found: the Hawking-Page-like phase transition in the grand canonical ensemble and the van der Waals-like phase transition in the canonical ensemble.  The phase structures of other thermodynamic systems such as black branes \cite{Lu:2010xt, Wu:2011yu, Lu:2012rm, Lu:2013nt}, boson stars \cite{Peng:2017squ} and hairy black holes \cite{Basu:2016srp, Peng:2017gss} in a cavity caught many interests.
Recently, the thermodynamics in AdS space and in a cavity were compared based on various black holes such as the nonlinear electrodynamics black hole \cite{Wang:2019kxp,Liang:2019dni}, the Gauss-Bonnet black hole \cite{Wang:2019urm} and the quintessence RN black hole \cite{Huang:2021iyf}. Besides, it shows that the thermodynamic phase space of a black hole can be extended by regarding the volume of the cavity as the thermodynamic volume \cite{Wang:2020hjw}. The phase transitions of the above black holes in a cavity demonstrate certain similarities and differences from the phase transitions of black holes in AdS space. However, these are all static black holes with dimensions no less than four. Consequently, it is nature to raise the questions that what will it be if the dimension is lower or if we introduce the angular momentum.

We start from considering a (2+1) dimensional black hole, namely BTZ black hole, which was first found by Banados, Teitelboim and Zanellit \cite{Banados:1992wn}. The BTZ black hole shows properties similar to a conventional (3+1) dimensional black hole. For example, a rotating BTZ black hole has both an inner and an outer horizon like the Kerr black hole, it has the ``surface area" proportional entropy and it is fully characterized by the ADM mass, angular momentum and the electric charge \cite{Banados:1992wn, Banados:1992gq}. The black hole solutions coupled different nonlinear electromagnetic fields were first studied in \cite{Hendi:2012zz}, which shows that the black holes are thermodynamically stable in  a canonical ensemble. The thermodynamic geometry for BTZ black holes was also investigated intensively under different boundary conditions \cite{Sarkar:2006tg, Wei:2009zzf, Akbar:2011qw, Ghosh:2020kba}. 

From the perspective of the phase transition, the BTZ black hole has also attracted many attentions. It shows that in three dimensional gravity, the phase transition between thermal AdS space and the black hole is also possible \cite{Birmingham:2002ph, Barbon:2003aq, Solodukhin:2005qy, Barbon:2003rou, Kurita:2004yn, Kleban:2004rx}. In three-dimensional gravity, there exists two distinct solutions, the BTZ black hole for $M\ge0$ and the thermal soliton for global AdS$_3$ with $M=-1$, i.e., the thermal AdS$_3$ space \cite{Kleban:2004rx, Horowitz:1998ha, Surya:2001vj}. However, it was pointed out that the phase transition between thermal AdS$_3$ space and the massive BTZ black hole is discontinues rather than of the Hawking-Page since there is a mass gap between thermal AdS$_3$ space and the BTZ black hole \cite{Myung:2005ee}. Thus Myung introduced the mass of  conical singularities and then verified that the phase transition could be possible through the off-shell approach \cite{Myung:2006sq}. Later on, the phase transition was revisited and it was found that the continuous off-shell free energy describing tunneling effect can be realized through non-equilibrium solitons \cite{Eune:2013qs}. It was not long ago, authors in \cite{McGough:2016lol} proposed that the $T\bar{T}$ deformed $\text{CFT}_2$ locates at the finite radial position of $\text{AdS}_3$, which further promotes us to investigate the (2+1) dimensional black hole in a cavity.

The structures of this paper are as follows: In section \ref{BHST}, we obtain the free energy by means of the Euclidean action for three types of BTZ black holes in a cavity: a static and neutral black hole, a rotating black hole and a charged black hole. As expected, the black holes in cavities are thermodynamically stable, so we turn to consider the phase transitions between thermal AdS$_3$ space, the black hole and the boundary state with the minimal free energy in section \ref{PT}. We fully study the phase transition in a grand canonical ensemble and exhibit the phase diagrams. The final discussions are presented in section \ref{CD}. The locally thermodynamic stability analysis of the rotating BTZ black hole and the charged BTZ black hole in a cavity is performed in appendix \ref{PHCP}. Furthermore, we present the phase diagrams of BTZ black holes without the cavity in appendix \ref{ptbbgc} to make a comparison with the case of the cavity existing.

\section{Black Hole Solutions and Thermodynamics}\label{BHST}

In this section, we briefly review the derivation of the black hole solutions from the Lagrangian formula. By imposing the temperature-fixed boundary condition on the wall of the cavity and evaluating the Euclidean action, we obtain the corrective thermodynamic quantities of the system in a cavity. Our discussions will be divided into three parts including a static and neutral BTZ black hole, a rotating BTZ black hole and a charged BTZ black hole. The concrete discussions are as follows.

\subsection{Static and Neutral BTZ Black Hole}

We consider the action of a (2+1) dimensional black hole which consists of a bulk term and a boundary term \cite{Banados:1992wn}
\begin{equation}
	\mathcal{S}=\frac{1}{2\pi}\int_{\mathcal{M}}d^3x\sqrt{-g}(R-2\Lambda)+\frac{1}{\pi}\int_{\partial\mathcal{M}}d^2x\sqrt{-h}(K-K_0),
\end{equation}
where $\Lambda$ is the three dimensional cosmological constant, $h$ is the determinant of the induced metric on the boundary $\partial\mathcal{M}$, $K$ is the extrinsic curvature on the boundary and $K_0$ is a counter term to avoid the divergence of the action.  Varying the action gives the vaccum field equation in three\textcolor{blue}{-}dimensional version
\begin{equation}
	R_{\alpha\beta}-\frac{1}{2}Rg_{\alpha\beta}-\frac{1}{l^2}g_{\alpha\beta}=0.\label{emotion}
\end{equation}
The metric for the static symmetric black hole has the following ansatz
\begin{equation}
	ds^2=-f(r)dt^2+\frac{1}{f(r)}dr^2+r^2 d\phi^2.\label{nnmetric1}
\end{equation}
Further solving the field equation with the given ansatz gives
\begin{equation}
	f(r)=-M+\frac{r^2}{l^2},\label{nnmetric2}
\end{equation}
where $M$ is the ADM mass and $l$ is the AdS radius, which is related to the cosmological constant as $\Lambda=-1/l^2$.

We now study the thermodynamics of the black hole in a cavity. It shows that the statistical mechanical partition function can be related to the on-shell  Euclidean action in the semi-classical approximation \cite{York:1986it}
\begin{equation}
	\mathcal{Z}\simeq e^{-\mathcal{S}_{E}},
\end{equation}
where the Euclidean action $\mathcal{S}_{E}$ is obtained by the analytic continuation of the action and so is the Euclidean time
\begin{equation}
	\begin{aligned}
		S&=i\mathcal{S}_{E},\\
		\tau&=it.
	\end{aligned}
\end{equation}
According to statistical mechanics, the relation between the free energy and the partition function is
\begin{equation}
	F=-T\ln{\mathcal{Z}}=T\mathcal{S}_{E}.
\end{equation}
In addition, to obtain the expression of temperature, we impose the boundary condition of the fixed temperature on the wall of the cavity by Euclidean time \cite{Wang:2019kxp}
\begin{equation}
	\int d\tau=\frac{1}{T\sqrt{f(r_B)}},\label{boundarycondition}
\end{equation}
On the other hand, the period of $\tau$ is given by the reciprocal of Hawking temperature $1/T_{h}$, which implies 
\begin{equation}
	T=\frac{T_h}{\sqrt{f(r_B)}}.\label{temperature}
\end{equation}
The entropy of the BTZ black hole is given by \cite{Banados:1992wn}
\begin{equation}
	S=4\pi r_{+}.\label{entropy}
\end{equation}
The concrete expression for the temperature of a static and neutral BTZ black hole in a cavity according to Eq. (\ref{nnmetric2}) and Eq. (\ref{temperature}) is
\begin{equation}
	T=\frac{1}{4\pi\sqrt{f(r_B  )}}\left(\frac{2r_+}{l^2}\right),\label{t1}
\end{equation}
where $f(r_B)=-r_+^2/l^2+r_B^2/l^2$. For the static symmetrically metric, we evaluate the Euclidean action related to the metric (\ref{nnmetric1}) and (\ref{nnmetric2}) in a cavity
\begin{equation}
	\begin{aligned}
		\mathcal{S}_{E}&=-\frac{1}{T\sqrt{f(r_B)}}\int^{r_B}_{r_+}dr\left(-2f'(r)-rf''(r)+\frac{2r}{l^2}\right)-\frac{2}{T\sqrt{f(r_B)}}\left(f(r_B)+\frac{r_B^2}{l^2}-\frac{r_B\sqrt{f(r_B)}}{l}\right)\\
		&=-\frac{2r_+^2}{Tl^2\sqrt{f(r_B)}}-\frac{2\sqrt{f(r_B)}}{T}+\frac{2r_B}{Tl}.
	\end{aligned}
\end{equation}
Thus, the Helmholtz free energy is
\begin{equation}
	\begin{aligned}
		F=T\mathcal{S}_{E}&=-\frac{2r_+^2}{l^2\sqrt{f(r_B)}}-2\sqrt{f(r_B)}+\frac{2r_B}{l}.\label{fren1}
	\end{aligned}
\end{equation}
The internal energy of the black hole in a cavity is obtained through the thermodynamic relation
\begin{equation}
	E=F+TS=-2\sqrt{f(r_B)}+\frac{2r_B}{l}.\label{nenergy}
\end{equation}

The heat capacity will be helpful to investigate the thermodynamic stability of the black hole. Here, it shows straightforwardly that the heat capacity at the constant cavity radius reads
\begin{equation}
	C_{r_B}=T\left(\frac{\partial S}{\partial T}\right)_{r_B}=\frac{8\pi^2 l T}{r_B^2}\left(r_B^2-r_+^2\right)^{3/2}>0,
\end{equation}
where we have used $r_+< r_B$ and $T>0$ to obtain the inequality. This result implies that for a given temperature and a given cavity radius, there is only one branch of black hole and it is thermodynamically stable.

\subsection{Rotating BTZ Black Hole}

The metric of a (2+1) dimensional stationary rotating black hole is given by \cite{Banados:1992wn}
\begin{equation}
	ds^2=-f(r)dt^2+\frac{1}{f(r)}dr^2+r^2\left(h(r)dt+d\phi\right)^2,
\end{equation}
where $h(r)$ is the shift function. Solving the field equation yields
\begin{equation}
	\begin{aligned}
		h(r)&=-\frac{J}{2r^2},\\
		f(r)&=-M+\frac{r^2}{l^2}+\frac{J^2}{4r^2}.\label{jmetric}
	\end{aligned}
\end{equation}
where $J$ is interpreted as the angular momentum.

According to Eq. (\ref{temperature}) and Eq. (\ref{jmetric}), the temperature of the rotating BTZ black hole in a cavity is given by
\begin{equation}
	T=\frac{1}{4\pi \sqrt{f(r_B)}}\left(\frac{2r_+}{l^2}-\frac{J^2}{2r_+^3}\right),\label{asd1}
\end{equation}
where $f(r_B)=-\frac{r_+^2}{l^2}-\frac{J^2}{4r_+^2}+\frac{r_B^2}{l^2}+\frac{J^2}{4r_B^2}$. Evaluating the Euclidean action for the rotating black hole gives the Gibbs free energy
\begin{equation}
	\begin{aligned}
	G=-\frac{2r_+^2}{l^2\sqrt{f(r_B)}}+\frac{J^2}{2r_B^2\sqrt{f(r_B)}}-2\sqrt{f(r_B)}+\frac{2r_B}{l},\label{1omega}
	\end{aligned}
\end{equation}
where we have used Eq. (\ref{asd1}). For the rotating black hole, it is reasonable to interpret the angular momentum and its conjugated quantity as the volume and the negative pressure respectively. We define the angular velocity, i.e., the conjugated quantity of angular momentum as
\begin{equation}
	\omega\equiv-\frac{J}{2r_B^2\sqrt{f(r_B)}}+\frac{J}{2r_+^2\sqrt{f(r_B)}}.\label{asd2}
\end{equation}
The Helmholtz free energy thus can be given by
\begin{equation}
	F=G+\omega J=-2\sqrt{f(r_B)}+\frac{2r_B}{l}-\frac{1}{\sqrt{f(r_B)}}\left(\frac{2r_+^2}{l^2}-\frac{J^2}{2r_+^2}\right).\label{hfree2}
\end{equation}
The internal energy is
\begin{equation}
	E=F+TS=-2\sqrt{f(r_B)}+\frac{2r_B}{l}.\label{abs4}
\end{equation}
It is easy to verify the following thermodynamic relations
\begin{equation}
	T=\frac{\partial E}{\partial S},\text{ }\omega=\frac{\partial E}{\partial J}.
\end{equation}
We then establish the first law of thermodynamics from those equations and quantities above
\begin{equation}
	dE=TdS+\omega dJ.
\end{equation}

The quantity $f(r_B)$ that appears under the root sign should be positive to keep the quantities physically meaningful
\begin{equation}
	f(r_B)=-\frac{r_+^2}{l^2}-\frac{J^2}{4r_+^2}+\frac{r_B^2}{l^2}+\frac{J^2}{4r_B^2}> 0\Longrightarrow J< \frac{2r_+r_B}{l}.\label{consJ}
\end{equation}
Besides, the temperature of the rotating black hole should be non-negative, and thus we have
\begin{equation}
	J\le \frac{2 r_+^2}{l},\label{constraintJ}
\end{equation}
which gives a stricter constraint than the cavity does since $r_+<r_B$. Moreover, this constraint equation can be used to prove that the heat capacity for the rotating black hole at the constant cavity radius and angular momentum is non-negative
\begin{equation}
	C_{r_B,J}=T\left(\frac{\partial S}{\partial T}\right)_{r_B,J}\ge0\label{hcj}
\end{equation}
and so is the heat capacity at the constant cavity radius and the angular velocity
\begin{equation}
	C_{r_B,\omega}=T\left(\frac{\partial S}{\partial T}\right)_{r_B,\omega}\ge0.\label{hcj2}
\end{equation}
The positive heat capacity is sufficient to deduce that the non-extreme rotating black hole in a cavity is locally thermodynamically stable in a canonical ensemble. In a grand canonical ensemble, we also find that the non-extreme rotating black hole in a cavity is locally thermodynamically stable by analyzing the Hessian matrix of the internal energy. The detailed proof is presented in appendix \ref{PHCP}.

\subsection{Charged BTZ Black Hole}

The action of the black hole coupled to electromagnetism is given by \cite{Martinez:1999qi}
\begin{equation}
	\begin{aligned}
		\mathcal{S}=\frac{1}{2\pi}\int_{\mathcal{M}}d^3x\sqrt{-g}&(R-2\Lambda)+\frac{1}{\pi}\int_{\partial\mathcal{M}}d^2x\sqrt{-h}(K-K_0)\\
		&-\frac{1}{4}\int_{\mathcal{M}}d^3x\sqrt{-g}F^{\mu\nu}F_{\mu\nu}-\int_{\partial \mathcal{M}}d^2x\sqrt{-h}n_{\nu}F^{\mu\nu}A_{\mu},
	\end{aligned}
\end{equation}
where $F^{\mu\nu}$ is the electromagnetic field tensor, $n_{\nu}$ is the unit outward-pointing normal vector of $\partial \mathcal{M}$ and $A_{\mu}$ is the electromagnetic potential. Notice the last term in the action is to fix the charge on the boundary \cite{Braden:1990hw}. Varying the action with respect to $g_{\alpha\beta}$ and $A_{\mu}$ gives the equations of motion
\begin{equation}
	\begin{aligned}
		R_{\alpha\beta}-\frac{1}{2}Rg_{\alpha\beta}-\frac{1}{l^2}g_{\alpha\beta}&=\pi\left(F_{\alpha}{}^{\mu}F_{\beta\mu}-\frac{1}{4}g_{\alpha\beta}F^{\mu\nu}F_{\mu\nu}\right),\\
		\nabla_{\beta}F^{\alpha\beta}&=0.\label{em3}
	\end{aligned}
\end{equation}
We assume that the (2+1) dimensional static symmetrically solution has the following form
\begin{equation}
	ds^2=-f(r)dt^2+\frac{1}{f(r)}dr^2+r^2d\phi^2.
\end{equation}
The simplest case is, the electromagnetic field tensor has no components along the $\phi$ direction which ensures that the field is purely electric for an stationary observer. Here, we write down the metric straightforwardly \cite{Martinez:1999qi}
\begin{equation}
	f(r)=-M+\frac{r^2}{l^2}-\frac{Q^2}{2}\ln{\frac{r}{l}},\label{fcharged}
\end{equation}
where $Q$ is the electric charge of the black hole.

The temperature of the charged BTZ black hole in a cavity according to Eq. (\ref{temperature}) and Eq. (\ref{fcharged}) is given by
\begin{equation}
	T=\frac{1}{4\pi \sqrt{f(r_B)}}\left(\frac{2r_+}{l^2}-\frac{Q^2}{2r_+}\right),\label{temq}
\end{equation}
where $f(r_B)=-\frac{r_+^2}{l^2}+\frac{r_B^2}{l^2}+\frac{Q^2}{2}\ln{\frac{r_+}{r_B}}$. We evaluate the Helmholtz free energy through the Euclidean action
\begin{equation}
	\begin{aligned}
    	F=T\mathcal{S}_{E}&=-\frac{1}{\sqrt{f(r_B)}}\left(\frac{2r_+^2}{l^2}-\frac{Q^2}{2}\right)-2\sqrt{f(r_B)}+\frac{2r_B}{l}.\label{Hefreen}
	\end{aligned}
\end{equation}
The internal energy thus can be written as
\begin{equation}
	E=F+TS=-2\sqrt{f(r_B)}+\frac{2r_B}{l}.
\end{equation}
We define the potential conjugated to the electric charge
\begin{equation}
	\Phi\equiv\left(\frac{\partial E}{\partial Q}\right)=\frac{Q}{\sqrt{f(r_B)}}\ln{\frac{r_B}{r_+}},\label{phiq}
\end{equation}
which can be regarded as the negative pressure while the charge is regarded as the volume in the thermodynamic system. Therefore, the Gibbs free energy is given by
\begin{equation}
	G= F-\Phi Q=-2\sqrt{f(r_B)}+\frac{2r_B}{l}-\frac{2r_+^2}{l^2\sqrt{f(r_B)}}+\frac{Q^2}{2\sqrt{f(r_B)}}-\frac{Q^2}{\sqrt{f(r_B)}}\ln{\frac{r_B}{r_+}}.\label{grapotq}
\end{equation}
It is straightforward to verify that the temperature can be expressed as the differentiation of the thermal energy $E$ respect to entropy $S$
\begin{equation}
	T=\frac{\partial E}{\partial S}.\label{temaa}
\end{equation}
From Eq. (\ref{temq}) to Eq. (\ref{temaa}), we establish the first law of thermodynamics
\begin{equation}
	dE=TdS+\Phi dQ.
\end{equation}

Similarly, the charge of the black hole in a cavity is supposed to satisfy
\begin{equation}
	f(r_B)=-\frac{r_+^2}{l^2}+\frac{Q^2}{2}\ln{\frac{r_+}{l}}+\frac{r_B^2}{l^2}-\frac{Q^2}{2}\ln{\frac{r_B}{l}}> 0\Longrightarrow Q< \frac{1}{l}\sqrt{\frac{2(r_B^2-r_+^2)}{\ln (r_B/r_+)}}.\label{consQ}
\end{equation}
The constraint on the charge by requiring that the temperature is non-negative reads
\begin{equation}
	Q\le \frac{2r_+}{l},
\end{equation}
which can also be easily verified to be a stricter constraint. It shows that the heat capacity at the constant cavity radius and charge
\begin{equation}
	C_{r_B,Q}=T\left(\frac{\partial S}{\partial T}\right)_{r_B,Q}\ge0\label{qhcap}
\end{equation} 
and the heat capacity at the constant cavity radius and electric potential
\begin{equation}
	C_{r_B,\Phi}=T\left(\frac{\partial S}{\partial T}\right)_{r_B,\Phi}\ge0.\label{qhcap2}
\end{equation}
The positive heat capacity implies that the non-extreme charged BTZ black hole in a cavity in a canonical ensemble is locally thermodynamically stable. The detailed proof is presented in appendix \ref{PHCP}, where we also prove that the charged BTZ black hole in a cavity in a grand canonical ensemble is locally thermodynamically stable.

\section{Phase Transitions}\label{PT}

In the previous section, we have verified that for the three types of black holes, the heat capacities in the cavity are always non-negative, equivalently, there is no phase transition between different phases of black holes. However, there is the possibility that the phase transition occurs between thermal AdS$_3$ space and the BTZ black hole. Furthermore, due to the existence of the cavity, there might be additional candidates of the lowest free energy on the boundary of the physically allowed region \cite{Liang:2019dni,Huang:2021iyf}. Thus it is very intriguing to investigate the phase transitions of the BTZ black hole in a cavity. The metric of three\textcolor{blue}{-}dimensional thermal AdS space is obtained by replacing $M$ with $-1$ \cite{Banados:1992wn}
\begin{equation}
	ds^2=-\left(1+\frac{r^2}{l^2}\right)dt^2+\frac{1}{1+\frac{r^2}{l^2}}dr^2+r^2d\phi^2.
\end{equation}
Implementing a same procedure, the free energy of thermal AdS space can be calculated by means of the Euclidean action and we finally arrive at
\begin{equation}
	F_{AdS}=\frac{2r_B}{l}-2\sqrt{1+\frac{r_B^2}{l^2}}.\label{ffads}
\end{equation}
In order to simplify calculations, we rescale thermodynamic quantities by $r_B$
\begin{equation}
	\widetilde{r}_{+}\equiv \frac{r_+}{r_B},\text{ }\widetilde{l}\equiv \frac{l}{r_B},\text{ }\widetilde{J}\equiv\frac{J}{r_B},\text{ }\widetilde{\omega}\equiv\omega r_B,\text{ }\widetilde{Q}\equiv Q,\text{ }\widetilde{\Phi}\equiv\Phi,\text{ }\widetilde{T}\equiv T r_B,\text{ }\widetilde{F}\equiv F\text{ and }\text{ }\widetilde{G}\equiv G,\label{rescale}
\end{equation}
which is actually equivalent to setting $r_B=1$. Without loss of generality, we only consider the case when $\widetilde{\omega}$ and $\widetilde{\Phi}$ are positive. In the following part, we discuss the phase transitions of the three types of black holes, respectively.

\subsection{Static and Neutral BTZ Black Hole}

The temperature (\ref{t1}) and free energy (\ref{fren1}) of the static and neutral black hole are both functions of the horizon radius and AdS radius: $\widetilde{T}(\widetilde{r}_+,\widetilde{l})$, $\widetilde{F}(\widetilde{r}_+,\widetilde{l})$. For the thermal AdS space, the free energy (\ref{ffads}) is only the function of the AdS radius: $\widetilde{F}_{AdS}(\widetilde{l})$. We solve the equation $\widetilde{F}(\widetilde{r}_+,\widetilde{l})=\widetilde{F}_{AdS}(\widetilde{l})$ and plug its solution $\widetilde{r}_+(\widetilde{l})$ into the expression of temperature (\ref{t1}), which yields the temperature where the phase transition occurs
\begin{equation}
	\widetilde{T}_{c}=\frac{1}{2\pi}.\label{tc}
\end{equation}
We can also write it as $T_c=1/2\pi r_B$, so that the phase transition temperature is only determined by the radius of the cavity. Interestingly, without a cavity, the the phase transition temperature $T_c=1/2\pi l$ \cite{Eune:2013qs}, which only relates to the AdS radius.

\subsection{Rotating BTZ Black Hole}

\begin{figure}[ptb]
	\begin{center}
		\subfigure{\includegraphics[width=0.45\textwidth]{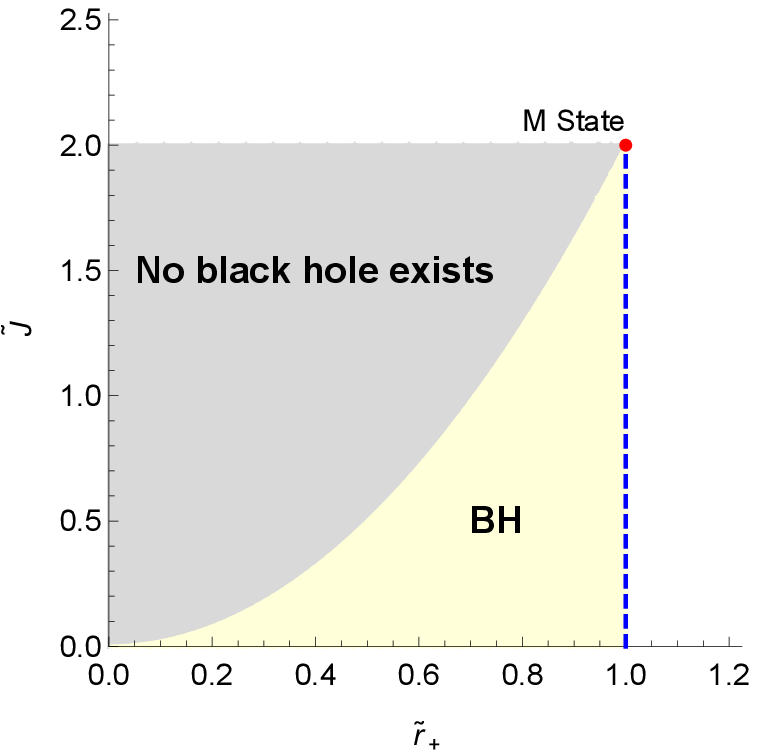}}
		\subfigure{\includegraphics[width=0.45\textwidth]{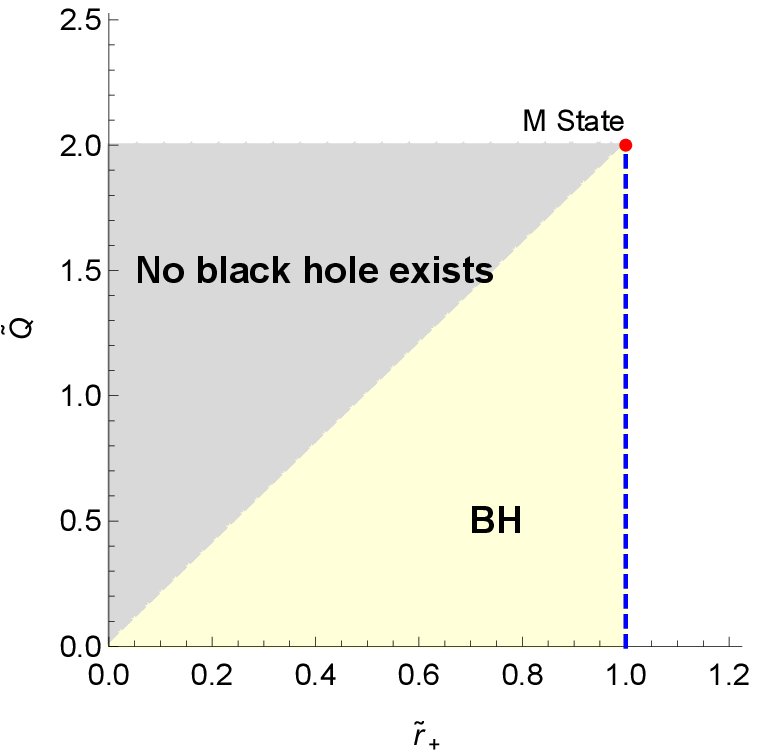}}
		\caption{Physically allowed regions for the rotating BTZ black hole in a cavity (the left panel) and the charged BTZ black hole in a cavity (the right panel) with $\widetilde{l}=1.00$. The blue dash line is the boundary of the physically allowed region, which is dubbed as BH. The M State which is marked in red has the minimal free energy on the boundary.}
		\label{AllowJ}
	\end{center}
\end{figure}

\begin{figure}[ptb]
	\begin{center}
		\subfigure{
			\includegraphics[width=0.45\textwidth]{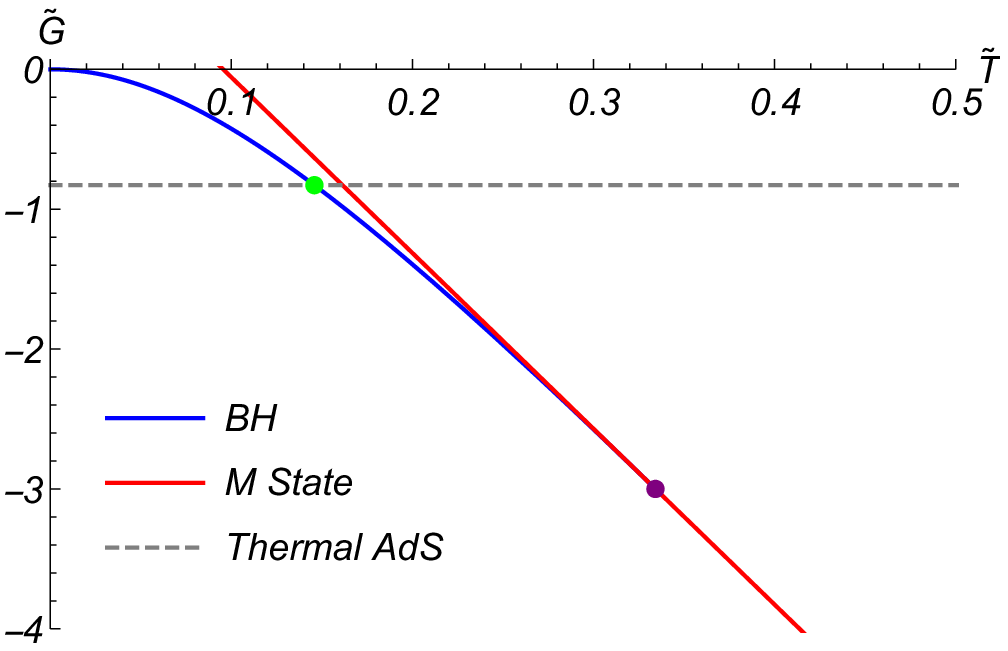}}
		\subfigure{
			\includegraphics[width=0.45\textwidth]{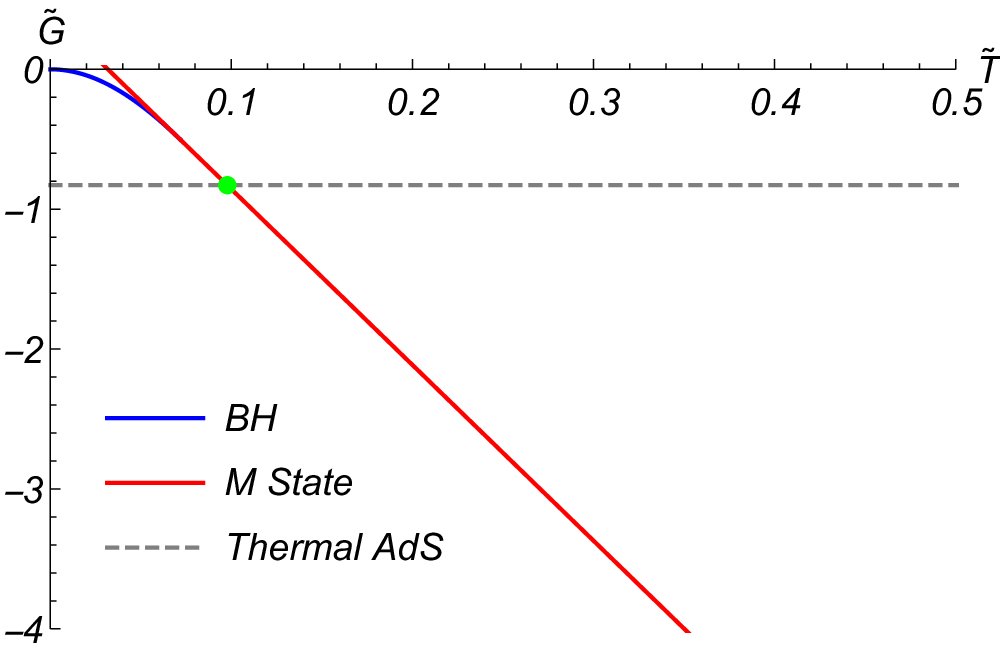}}
		\subfigure{
			\includegraphics[width=0.45\textwidth]{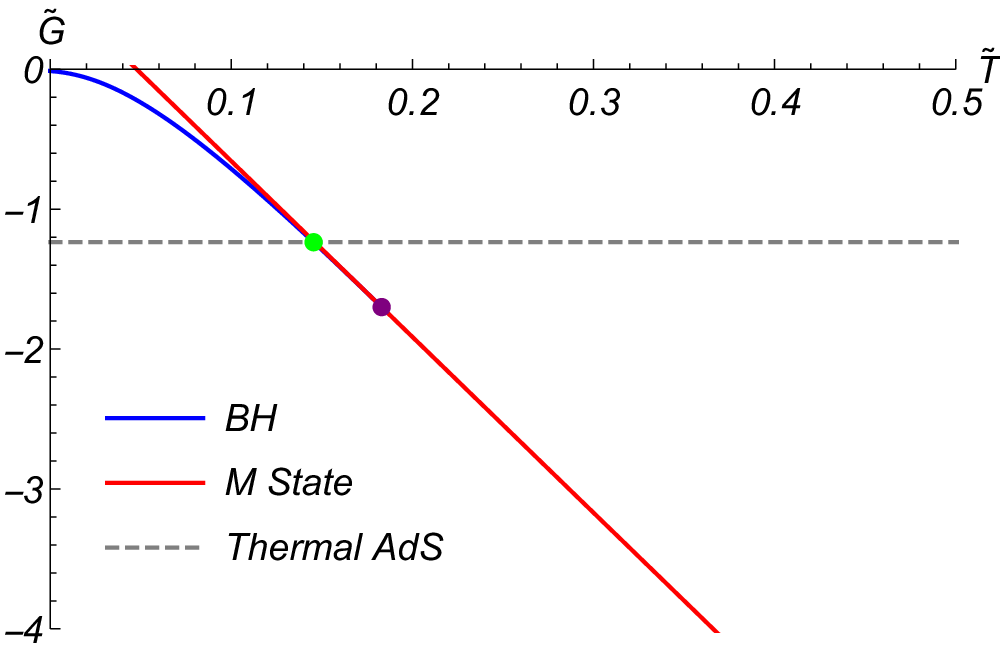}}
		\subfigure{
			\includegraphics[width=0.45\textwidth]{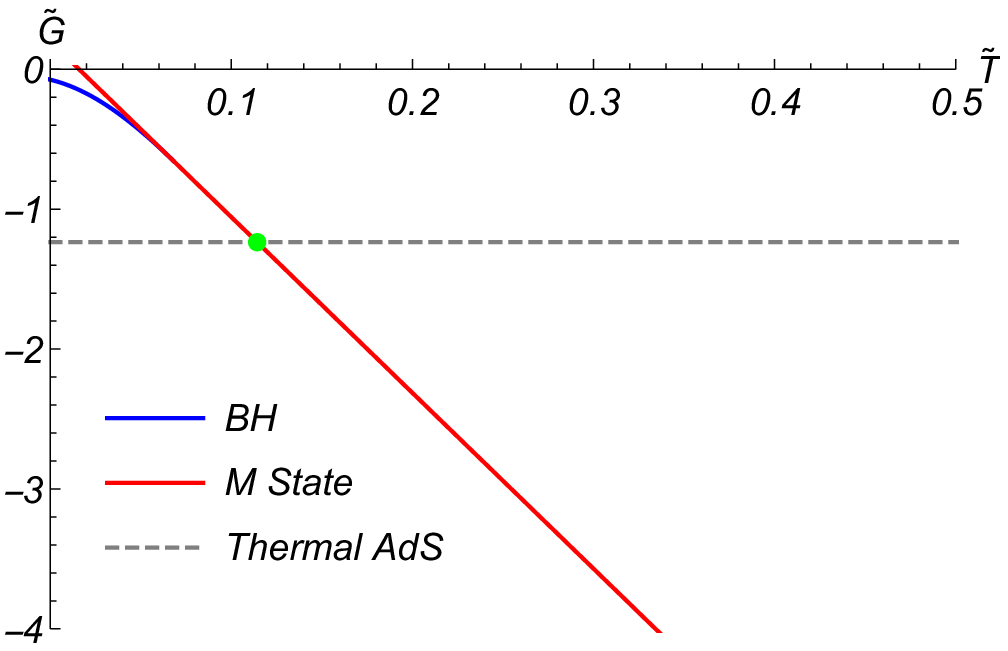}}
		\caption{Plots of Gibbs free energy against temperature of the rotating BTZ black hole in a cavity (upper two panels) and the charged BTZ black hole in a cavity (lower two panels). The green dot is where a first-order phase transition occurs and the purple dot is where a second-order phase transition occurs. \textbf{Upper Panel} The angular velocity $\widetilde{\omega}$ is fixed and the AdS radius $\widetilde{l}=1.00$. The values of the angular velocity $\widetilde{\omega}=0.40,\text{ }0.80$ from left to right. \textbf{Lower Panel} The potential $\widetilde{\Phi}$ is fixed and the AdS radius $\widetilde{l}=2.00$. The value of the potential $\widetilde{\Phi}=0.40,\text{ }0.80$ from left to right.}
		\label{GTJ}
	\end{center}
\end{figure}

\begin{figure}[ptb]
	\begin{center}
		\subfigure{
			\includegraphics[width=0.35\textwidth]{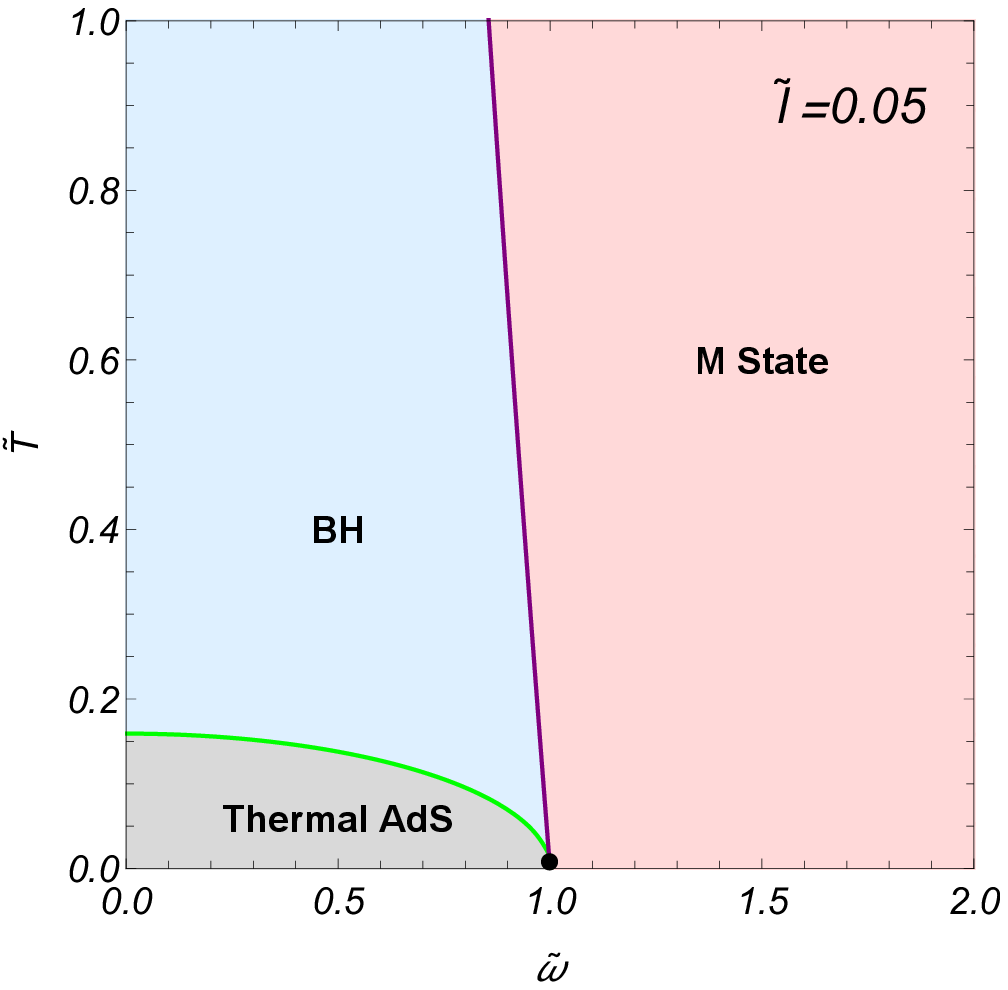}}
		\subfigure{
			\includegraphics[width=0.35\textwidth]{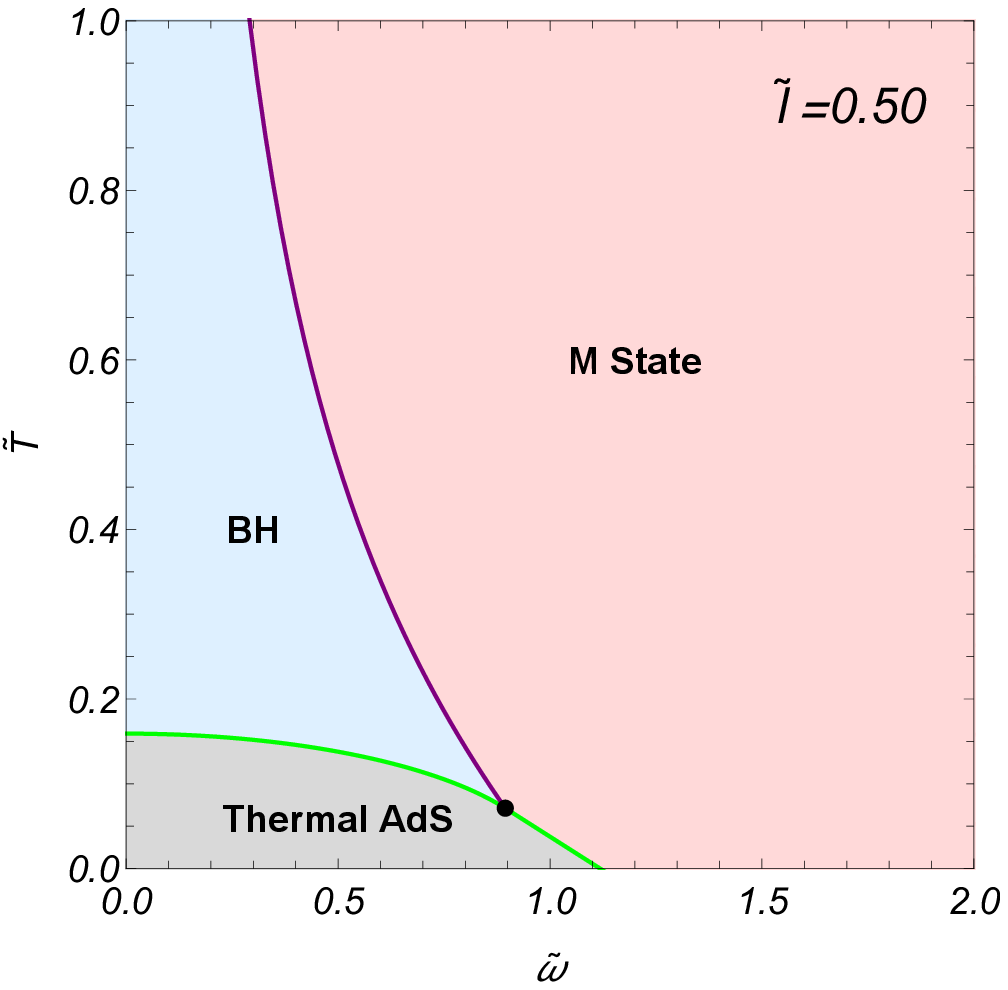}}
		
		\subfigure{
			\includegraphics[width=0.35\textwidth]{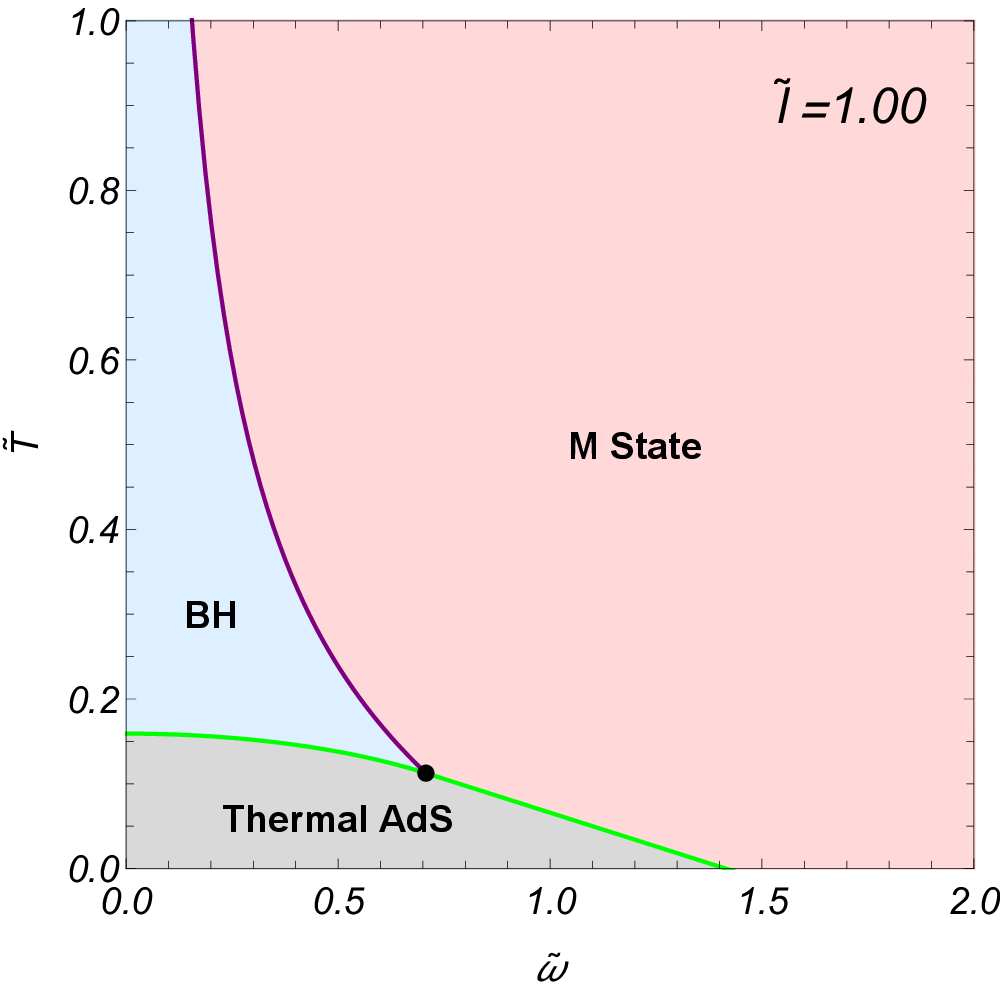}}
		
		\subfigure{
			\includegraphics[width=0.35\textwidth]{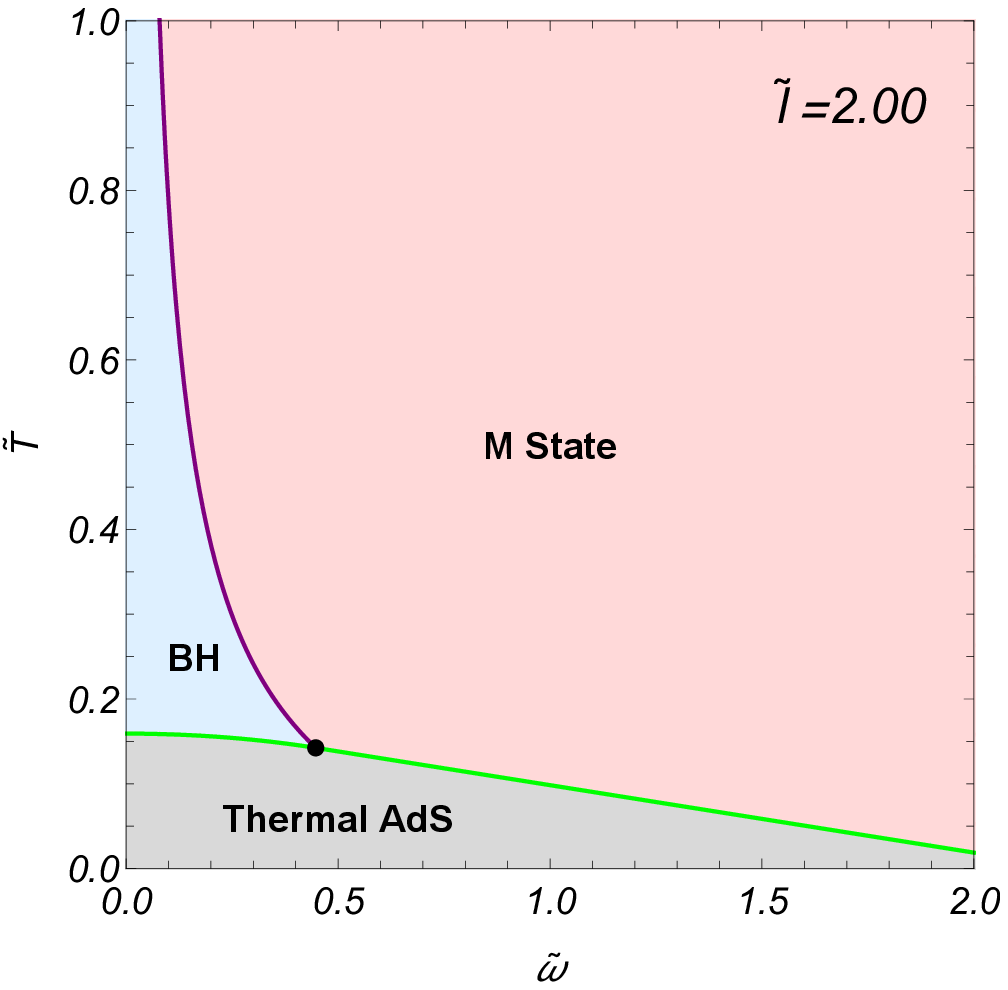}}
		\subfigure{
			\includegraphics[width=0.35\textwidth]{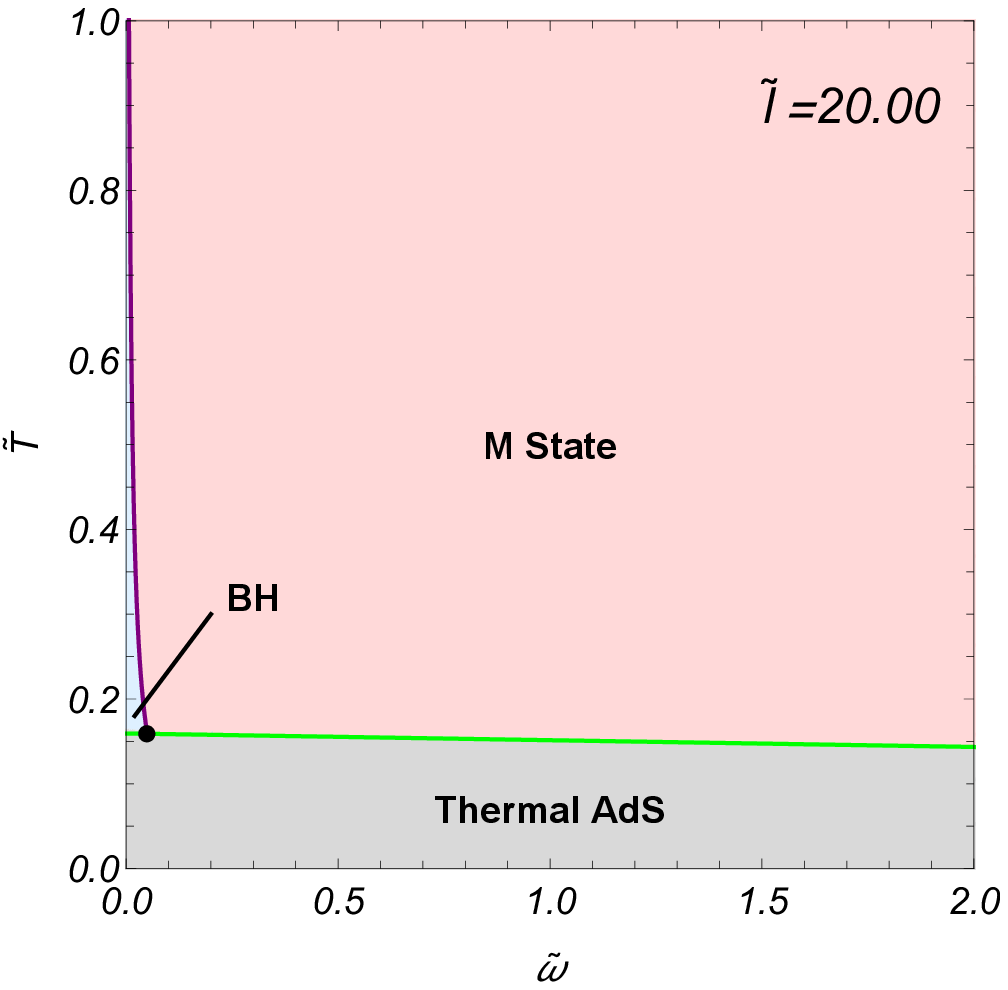}}
		\caption{Phase diagrams of the rotating BTZ black hole in a cavity. The AdS radius $\widetilde{l}=0.05,\text{ }0.50,\text{ }1.00,\text{ }2.00,\text{ }20.00$, respectively. The green curve is a first-order phase transition curve and the purple curve is a second-order phase transition curve. The black dot where these curves intersect is a triple point. }
		\label{RJ}
	\end{center}
\end{figure}

We consider the thermodynamic system in a grand canonical ensemble, which has the fixed temperature and angular velocity. The related thermodynamic quantities are $\widetilde{T}(\widetilde{r}_+,\widetilde{J},\widetilde{l})$, $\widetilde{G}(\widetilde{r}_+,\widetilde{J},\widetilde{l})$ and $\widetilde{\omega}(\widetilde{r}_+,\widetilde{J},\widetilde{l})$ in Eq. (\ref{asd1}), Eq. (\ref{1omega}) and Eq. (\ref{asd2}). $\widetilde{\omega}(\widetilde{r}_+,\widetilde{J},\widetilde{l})$ can be inverted to give the angular momentum $\widetilde{J}(\widetilde{r}_+,\widetilde{\omega},\widetilde{l})$, which can be inserted in $\widetilde{T}(\widetilde{r}_+,\widetilde{J},\widetilde{l})$ and $\widetilde{G}(\widetilde{r}_+,\widetilde{J},\widetilde{l})$ to yield $\widetilde{T}(\widetilde{r}_+,\widetilde{\omega},\widetilde{l})$ and $\widetilde{G}(\widetilde{r}_+,\widetilde{\omega},\widetilde{l})$. It turns out that our expressions for the temperature, angular velocity and free energy are reasonable when $r_+<r_B$. For the case that the event horizon merges with the cavity, we are supposed to discuss it separately. We start with the constraint of the physical quantities in a cavity from Eq. (\ref{constraintJ}). The physically allowed region of the rotating BTZ black hole is shown in the left panel of FIG. \ref{AllowJ}, where the region that our expressions are well defined is marked in yellow and dubbed as BH. The blue dashed line represents the case when $r_+=r_B$, that is, the black hole merges with the cavity. The Gibbs free energy of this state is given by
\begin{equation}
	\widetilde{G}=-4\pi \widetilde{T}+\frac{2}{\widetilde{l}}-\widetilde{\omega}\widetilde{J},
\end{equation}
where we have used the expression of the energy (\ref{abs4}). On this boundary we have $\partial{\widetilde{G}}/\partial{\widetilde{J}}=-\widetilde{\omega}<0$, which indicates that the lowest free energy point is located at the point with the maximal $\widetilde{J}$. The lowest free energy point of this boundary is marked in red and dubbed as M state. The Gibbs free energy of M State thus is
\begin{equation}
	\widetilde{G}_{M}=-4\pi\widetilde{T}+\frac{2}{\widetilde{l}}\left(1-\widetilde{\omega}\right).
\end{equation}
Plots of Gibbs free energy of BH, M state and Thermal AdS with respect to temperature are shown in the upper panel of FIG. \ref{GTJ}, in which we fix the parameter $\widetilde{l}=1.00$ and $\widetilde{\omega}=0.40\text{ and }0.80$ from left to right. For a small value of $\widetilde{\omega}$ at a low temperature, Thermal AdS is the stable phase. As temperature increases, a first order phase transition occurs from Thermal AdS space to BH and BH becomes the stable phase. As temperature further increases, a second-order phase transition (this can be easily verified according to the behavior of free energy with respect to temperature) occurs from BH to M State, and M State then becomes the stable phase. For a large $\widetilde{\omega}$, there is only a first-order phase transition between Thermal AdS and M State. The phase diagrams of the rotating BTZ black hole in a grand canonical ensemble are exhibited in FIG. \ref{RJ}, in which we find a triple point always exists among the three phases. Moreover, as the AdS radius $\widetilde{l}$ increases, the area of Thermal AdS in phase space apparently increases.

\subsection{Charged BTZ Black Hole}

\begin{figure}[ptb]
	\begin{center}
		\subfigure{
			\includegraphics[width=0.35\textwidth]{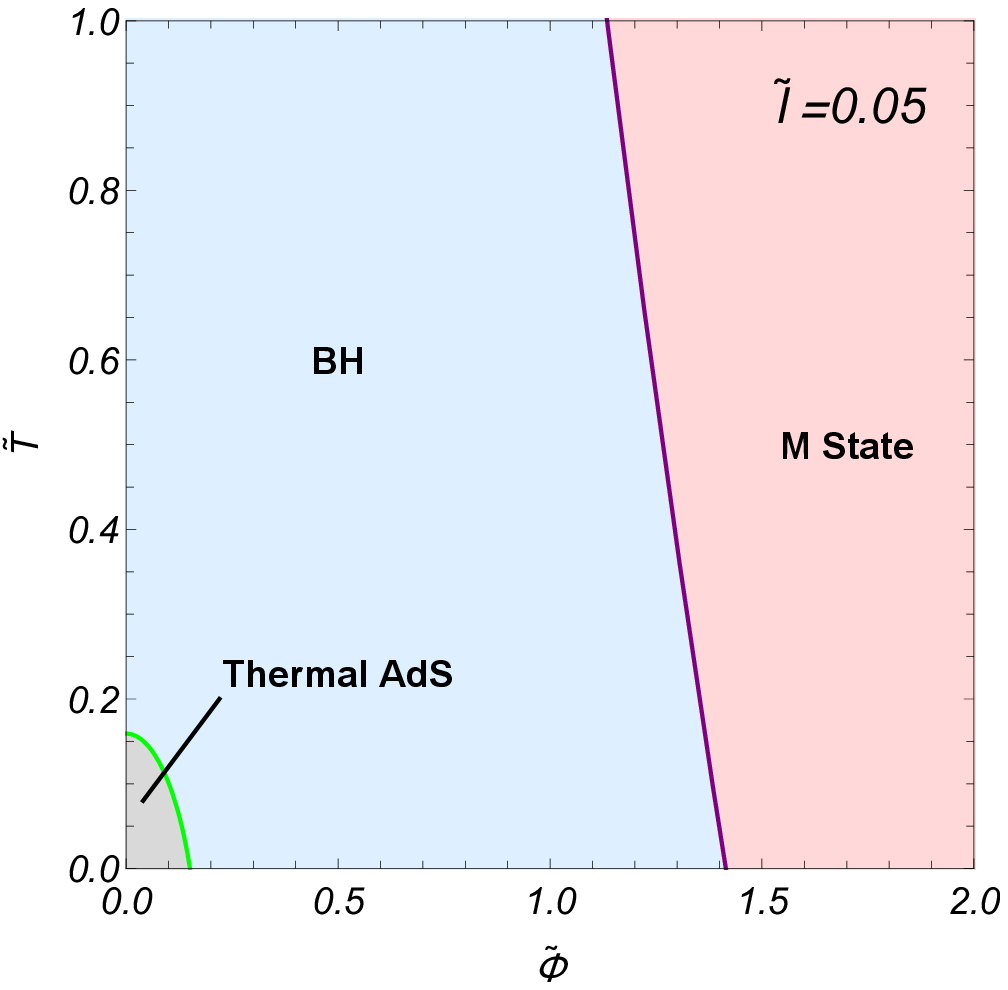}}
		\subfigure{
			\includegraphics[width=0.35\textwidth]{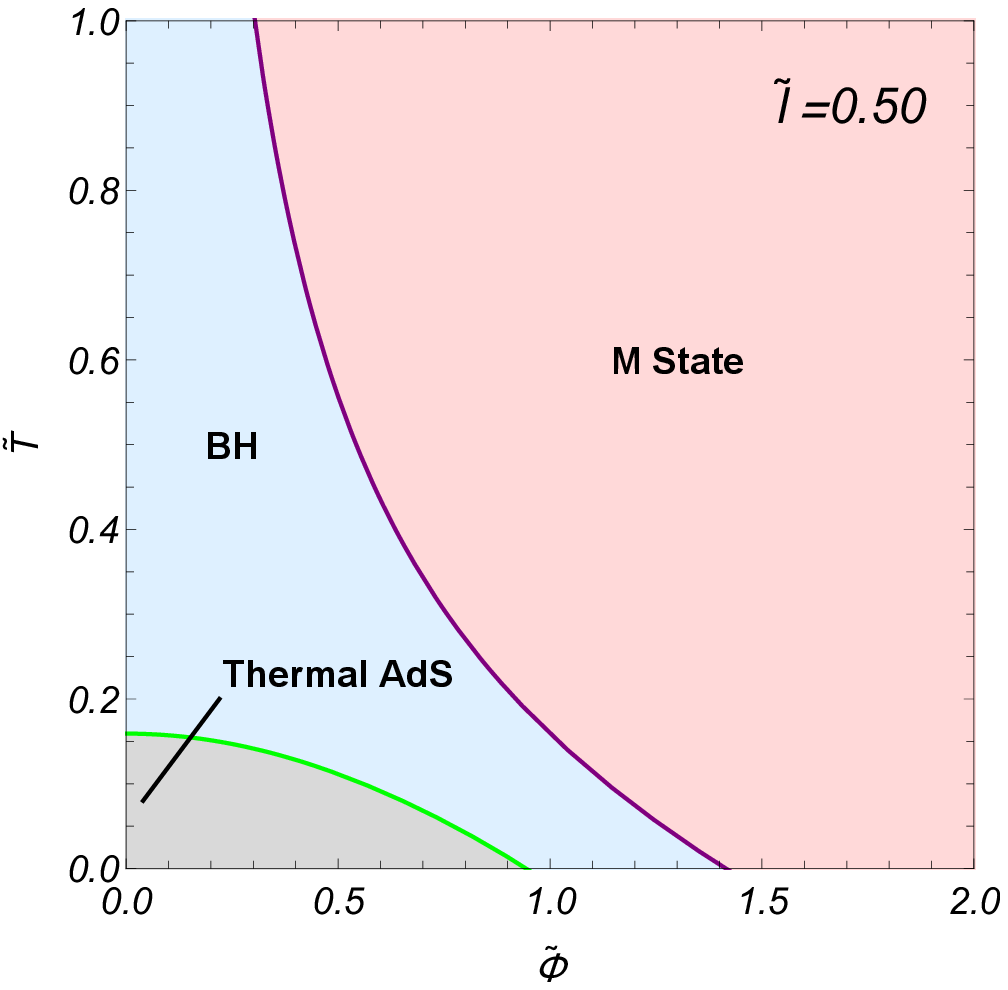}}
		
		\subfigure{
			\includegraphics[width=0.35\textwidth]{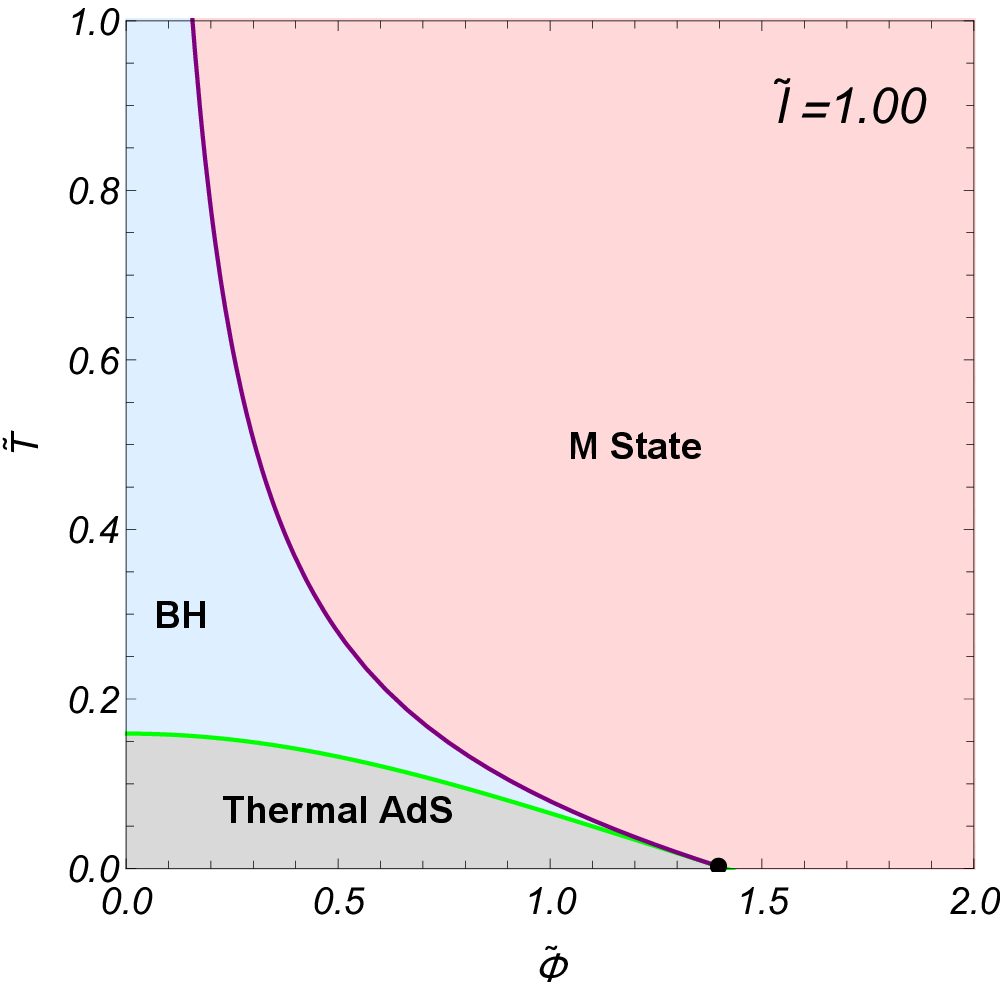}}
		
		\subfigure{
			\includegraphics[width=0.35\textwidth]{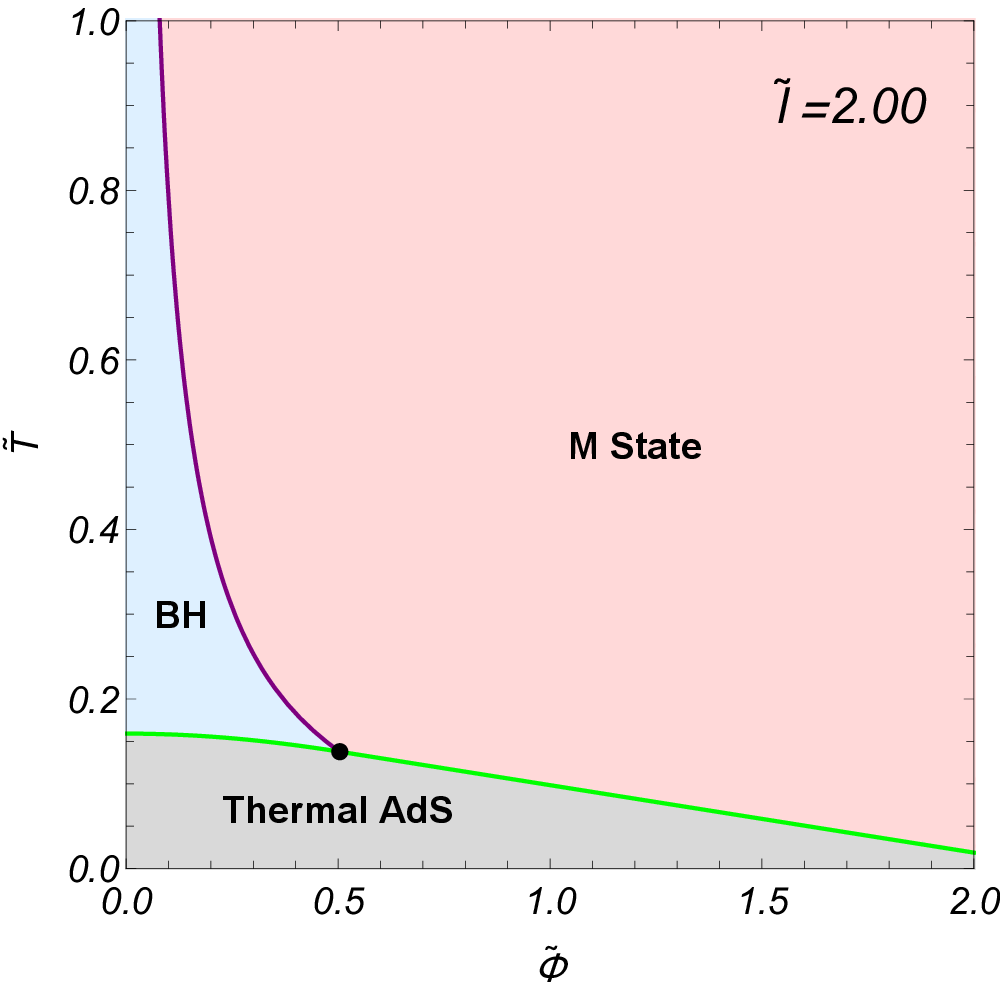}}
		\subfigure{
			\includegraphics[width=0.35\textwidth]{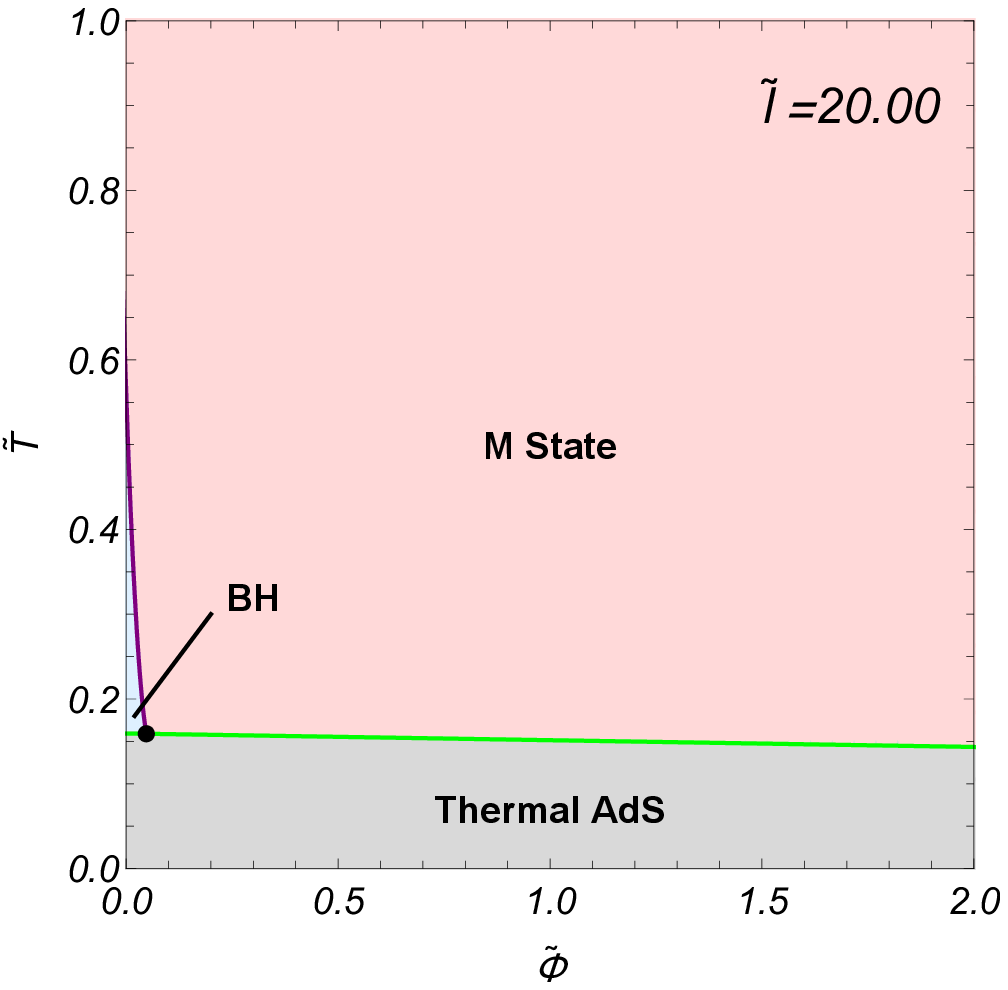}}
		\caption{Phase diagrams of the charged BTZ black hole in a cavity. The AdS radius $\widetilde{l}=0.05,\text{ }0.50,\text{ }1.00,\text{ }2.00,\text{ }20.00$, respectively. The green curve is a first-order phase transition curve and the purple curve is a second-order phase transition curve. The black dot where these curves intersect is a triple point. }
		\label{RQ}
	\end{center}
\end{figure}

The analysis of the charged black hole in the grand canonical ensemble is similar to the rotating one. Through thermodynamic quantities $\widetilde{T}(\widetilde{r}_+,\widetilde{Q},\widetilde{l})$, $\widetilde{G}(\widetilde{r}_+,\widetilde{Q},\widetilde{l})$ and $\widetilde{\Phi}(\widetilde{r}_+,\widetilde{Q},\widetilde{l})$ in Eq. (\ref{temq}), Eq. (\ref{phiq}) and Eq. (\ref{grapotq}), we obtain $\widetilde{T}(\widetilde{r}_+,\widetilde{\Phi},\widetilde{l})$ and $\widetilde{G}(\widetilde{r}_+,\widetilde{\Phi},\widetilde{l})$. As mentioned before, these formulas are only valid in a specific region, which is shown in the right panel of FIG. \ref{AllowJ}. On the boundary where $r_+=r_B$, we have
\begin{equation}
	\widetilde{G}=-4\pi\widetilde{T}+\frac{2}{\widetilde{l}}-\widetilde{\Phi}\widetilde{Q}.
\end{equation}
The lowest free energy is also located at the point with the maximal $\widetilde{Q}$, which is dubbed as M State and marked in red. Therefore, the Gibbs free energy of M State is
\begin{equation}
	\widetilde{G}_{M}=-4\pi\widetilde{T}+\frac{2}{\widetilde{l}}\left(1-\widetilde{\Phi}\right).
\end{equation}
Plots of Gibbs free energy against the temperature are displayed in the lower panel of FIG. \ref{GTJ}, where the parameters AdS radius $\widetilde{l}=2.00$ and the potential $\widetilde{\Phi}=0.40,\text{ }0.80$ from left to right. For a small value of $\widetilde{\Phi}$, as temperature increases, a first-order phase transition occurs from Thermal AdS to BH. As temperature further increases, a second-order phase transition occurs from BH to M State. For a large value of $\widetilde{\Phi}$, only a first-order phase transition occurs from Thermal AdS to M State. Phase diagrams are shown in FIG. \ref{RQ}, where we also find a triple point that is similar to the rotating black hole but only for a large enough AdS radius. As AdS radius $\widetilde{l}$ increases, the area of Thermal AdS in phase space also increases.

\section{Conclusion and Discussion}\label{CD}

Beginning with the gravitational action, we derived the spacetime solution of a static and neutral BTZ black hole. Through imposing the boundary condition, i.e., assigning the period of Euclidean time to $1/T\sqrt{f(r_B)}$, we obtained the free energy of the black hole. Then by computing the Euclidean action and free energy, we constructed the first law of thermodynamics of the system in a cavity. Analogously, the thermodynamic quantities of the rotating black hole and the electromagnetic field coupled black hole solutions were obtained. We proved that the heat capacities of these black holes in a canonical or grand canonical ensemble are always non-negative in appendix \ref{PHCP}, where we also proved that the thermodynamic systems of the non-extreme rotating black hole and the charged black hole are locally stable in a grand canonical ensemble.

For a static and neutral BTZ black hole in a cavity, we found that there is a critical temperature (the temperature that the phase transition occurs). Below the critical temperature, thermal AdS space is a stable state and above the critical temperature, the black hole is the stable state, which is similar to the case of no cavity. To our surprise, the critical temperature is only related to the cavity radius rather than AdS radius. Once taking angular momentum and charge into considerations, the phase transitions will show some intriguing properties. For a rotating BTZ black hole, we found an extra second-order phase transition between BH and the black hole-cavity merger state, which is dubbed as M State. Phase diagrams of various AdS radius $\widetilde{l}$ were exhibited in FIG. \ref{RJ}. A triple point was found to be located at the interaction of the first-order phase transition curve and the second-order phase transition curve. For a charged BTZ black hole, a similar second-order phase transition between BH and M State was also found. The phase diagrams were shown in FIG. {\ref{RQ}}. Unlike the rotating BTZ black hole, the triple point of the charged BTZ black hole system only exists for a large AdS radius $\widetilde{l}$.

The phase transitions of BTZ black holes without cavity were discussed in appendix \ref{ptbbgc}. We found that the phase transitions of a cavity existing and without a cavity show some dissimilarities. On the one hand, the second-order phase transition only exists in a cavity but no for the cavity absenting. On the other hand, as the AdS radius increases, the area of Thermal AdS in phase space increases in a cavity. For the case that the cavity does not exist, the area of Thermal AdS always decreases with an increasing AdS radius.

\begin{acknowledgments}
	We are grateful to Peng Wang, Bo Ning, Hanwen Feng and Yihe Cao for useful discussions and valuable comments. This work is supported by NSFC (Grant No.11947408 and 12047573).
\end{acknowledgments}

\appendix

\section{Thermodynamic stability of BTZ black hole in a cavity}\label{PHCP}

In this appendix, we are about to prove that the four types of heat capacities $C_{r_B,J}$, $C_{r_B,\omega}$, $C_{r_B,Q}$, $C_{r_B,\Phi}$ are non-negative as well as to analyse the thermodynamic stability of the BTZ black hole in a cavity. Notice that the heat capacity can always be written in form
\begin{equation}
	C_{\text{fixed quantities}}=4\pi T\bigg/\left(\frac{\partial T}{\partial r_+}\right)_{\text{fixed quantities}},
\end{equation}
where the temperature $T$ we always assume to be non-negative, and hence the heat capacity has the same sign of $\partial T/\partial r_+$. To simplify the calculation, we only consider this term.

\subsection{$C_{r_B,J}\ge0$}

\begin{figure}[ptb]
	\begin{center}
		\subfigure{
			\includegraphics[width=0.4\textwidth]{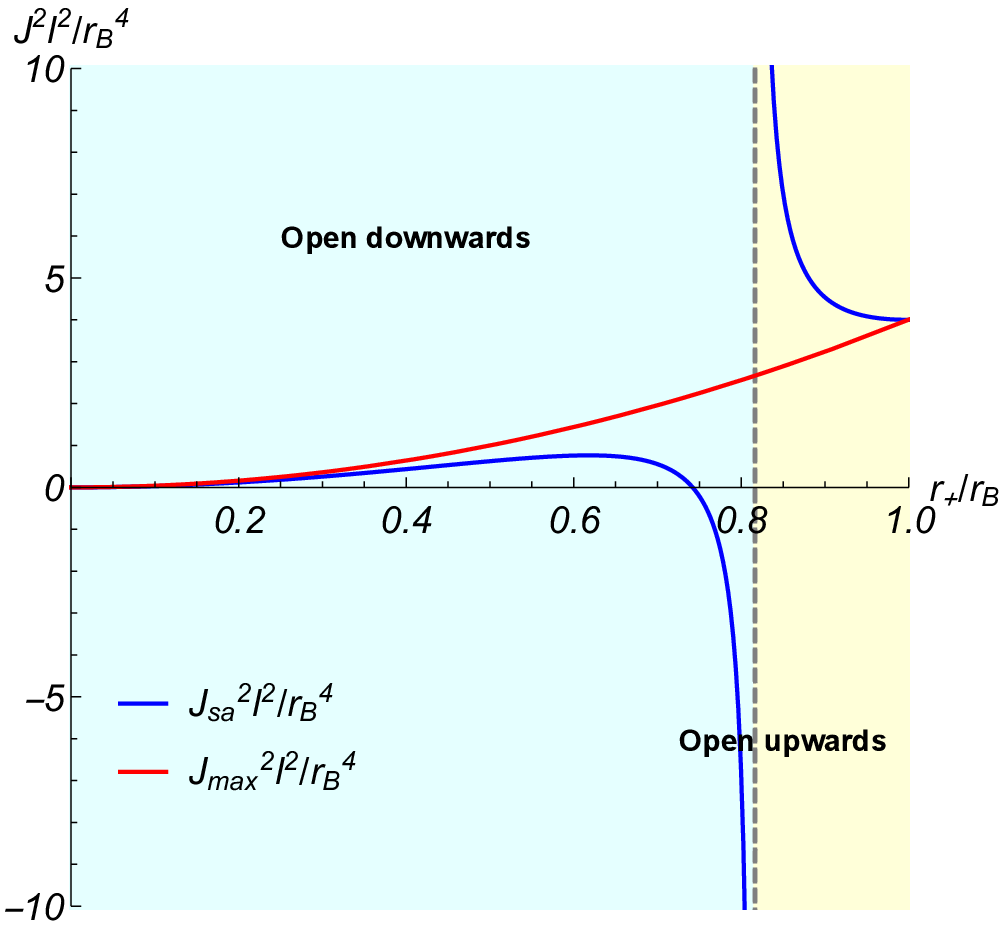}}
		\subfigure{
			\includegraphics[width=0.4\textwidth]{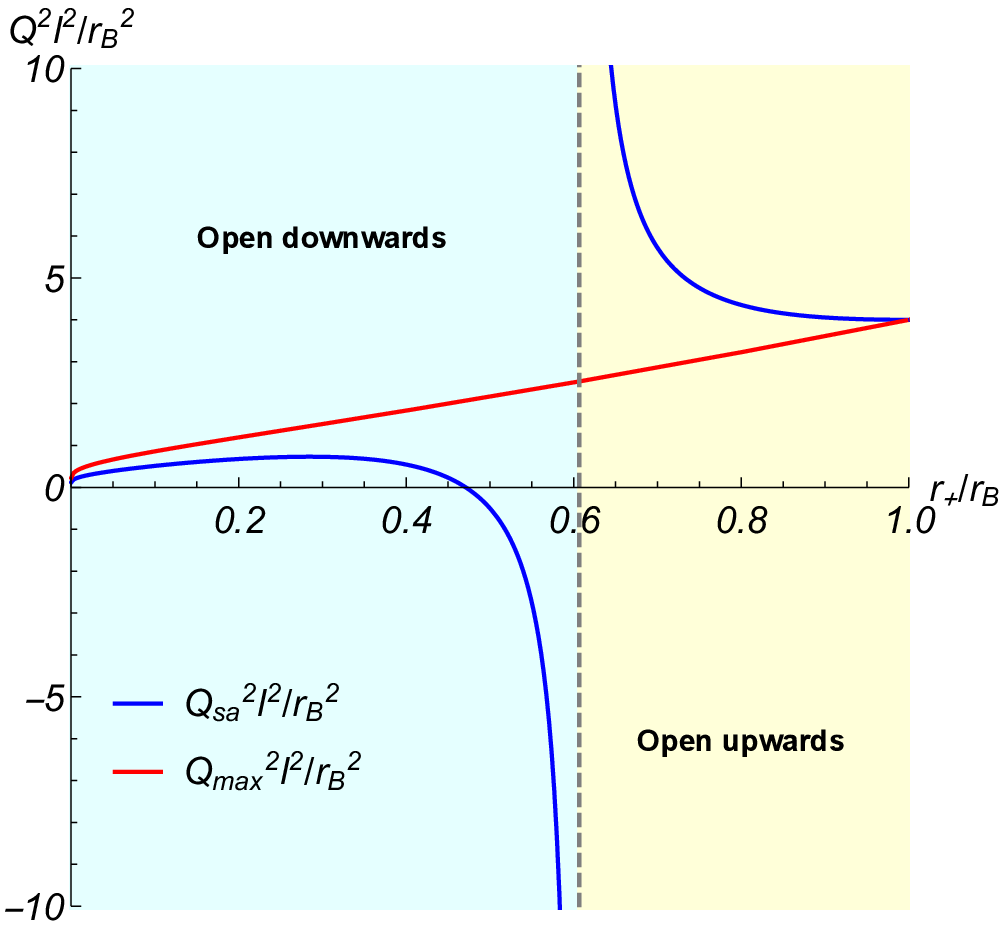}}
		\caption{\textbf{Left Panel} Plots of $J_{\text{sa}}^2 l^2/r_B^4$ and $J_{\text{max}}^2 l^2/r_B^4$ with respect to $r_+/r_B$. $g(Q^2)$ opens upwards in the yellow region and opens downwards in the blue region. $g(J^2)$ is a linear function on the gray dashed vertical line which is located on $r_+/r_B=\sqrt{6}/3\approx$0.82. \textbf{Right Panel} Plots of $Q_{\text{sa}}^2 l^2/r_B^2$ and $Q_{\text{max}}^2 l^2/r_B^2$ with respect to $r_+/r_B$. $k(Q^2)$ opens upwards in the yellow region and opens downwards in the blue region. $k(Q^2)$ is a linear function on the gray dashed vertical line which is located on $r_+/r_B=e^{-1/2}\approx$0.61.}
		\label{app}
	\end{center}
\end{figure}

According to the expression of temperature (\ref{asd1}), we have
\begin{equation}
	\frac{\partial T}{\partial r_+}=
	\frac{4 J^2 l^2 r_+^2 \left(-6 r_B^2 r_+^2+3 r_B^4+r_+^4\right)+J^4 l^4 \left(3 r_+^2-2 r_B^2\right)+16 r_B^4 r_+^6}{4 \pi  l^2 r_+^4 \left(r_+^2-r_B^2\right) \left(J^2 l^2-4 r_B^2 r_+^2\right) \sqrt{J^2 \left(\frac{1}{r_B^2}-\frac{1}{r_+^2}\right)+\frac{4 (r_B-r_+) (r_B+r_+)}{l^2}}}.
\end{equation}
It is easy to check that the denominator is positive since $r_+<r_B$ and $J\le 2r_+^2/l$. We define a new function equals to the numerator
\begin{equation}
	g(J^2)=4 J^2 l^2 r_+^2 \left(-6 r_B^2 r_+^2+3 r_B^4+r_+^4\right)+J^4 l^4 \left(3 r_+^2-2 r_B^2\right)+16 r_B^4 r_+^6,
\end{equation}
which is regarded as a quadratic function of $J^2$. It follows that the signs of $\partial T/\partial r_+$ and $g(J^2)$ are the same. On the endpoints of the interval $(0,J_{\text{max}}^2)$, we have
\begin{equation}
	\begin{aligned}
		g\left(0\right)&=16r_+^6r_B^4>0,\\
		g\left(J_{\text{max}}^2\right)&=64r_+^6\left(r_B^2-r_+^2\right)^2>0,
	\end{aligned}
\end{equation}
where $J_{\text{max}}=2r_+^2/l$ is the maximum angular momentum of a BTZ black hole. The symmetry axis of $g(J^2)$ is given by
\begin{equation}
	J_{\text{sa}}^2=\frac{2 r_+^2 \left(-6 r_B^2 r_+^2+3 r_B^4+r_+^4\right)}{l^2 \left(2 r_B^2-3 r_+^2\right)}.
\end{equation}
We plot the curves of $J_{\text{sa}}^2 l^2/r_B^4$ and $J_{\text{max}}^2 l^2/r_B^4$ with respect to $r_+/r_B$ in the left panel of FIG. \ref{app}. The characteristics of $g(J^2)$ on the interval $(0,J_{\text{max}}^2)$ are discussed in three cases:
\begin{itemize}
	\item If $r_+/r_B>\sqrt{6}/3$, $g(J^2)$ will open upwards and $J_{\text{max}}^2<J_{\text{sa}}^2$, which means that the values of $g(J^2)$ on the interval $(0,J_{\text{max}}^2)$ are positive.
	\item If $r_+/r_B=\sqrt{6}/3$, $g(J^2)$ will be a linear function about $J^2$, which means that the values of $g(J^2)$ on the interval $(0,J_{\text{max}}^2)$ are always positive.
	\item If $r_+/r_B<\sqrt{6}/3$, $g(J^2)$ will open downwards and
	$J_{\text{sa}}^2<J_{\text{max}}^2$, which also ensures that $g(J^2)$ on the interval $(0,J_{\text{max}}^2)$ is always positive.
\end{itemize}
Therefore, we infer that $C_{r_B,J}\ge0$.

\subsection{$C_{r_B,\omega}\ge0$}

We invert Eq. (\ref{asd2}) to obtain $J(r_+,r_B,\omega,l)$, which can be inserted into the expression of temperature (\ref{asd1}) to give $T(r_+,r_B,\omega,l)$
\begin{equation}
	T=\frac{r_+(1-r_B^2\omega^2)}{2\pi l}\left(r_B^2-r_+^2+r_B^2r_+^2\omega^2\right)^{-1/2}.
\end{equation}
Differentiating $T$ with respect to $r_+$ gives
\begin{equation}
	\frac{\partial T}{\partial r_+}=\frac{r_B^2(1-r_B^2\omega^2)}{2\pi l}\left(r_B^2-r_+^2+r_B^2r_+^2\omega^2\right)^{-3/2}=\frac{(2\pi l)^2 r_B^2 T^3}{r_+^3 (1-r_B^2\omega^2)^2}\ge0.
\end{equation}
Thus the heat capacity $C_{r_B,\omega}\ge0$.

\subsection{$C_{r_B,Q}\ge0$}

According to Eq. (\ref{temq}), we arrive at
\begin{equation}
	\frac{\partial T}{\partial r_+}=
	\frac{4 l^2 Q^2 \left(r_B^2-3 r_+^2\right)+2 \left(4 l^2 Q^2 r_+^2+l^4 Q^4\right) \ln{\frac{r_+}{r_B}}+l^4 Q^4+16 r_B^2 r_+^2}{8 \sqrt{2} \pi  l r_+^2 \left(l^2 Q^2 \ln{\frac{r_+}{r_B}}+2 (r_B-r_+) (r_B+r_+)\right)^{3/2}},
\end{equation}
of which the denominator is positive while the numerator is a quadratic function of $Q^2$, which is denoted as $k(Q^2)$. On the endpoints of the interval $(0,Q_{\text{max}}^2)$, we have
\begin{equation}
	\begin{aligned}
		k\left(0\right)&=16r_+^2r_B^2>0,\\
		k\left(Q_{\text{max}}^2\right)&=32r_+^2\left(r_B^2-r_+^2+2r_+^2\ln{\frac{r_+}{r_B}}\right)>0,
	\end{aligned}
\end{equation}
where $Q_{\text{max}}=2r_+/l$ is the maximal charge that a BTZ black hole could have. The symmetry axis of $k(Q^2)$ is
\begin{equation}
	Q_{\text{sa}}^2=-\frac{2}{l^2}\left(r_+^2+\frac{r_B^2-4r_+^2}{1+2\ln{\frac{r_+}{r_B}}}\right).
\end{equation}
We plot the curves of $Q_{\text{sa}}^2 l^2/r_B^2$ and $Q_{\text{max}}^2 l^2/r_B^2$ with respect to $r_+/r_B$ in the right panel of FIG. \ref{app}. Further discussions are as follows:
\begin{itemize}
	\item If $r_+/r_B>e^{-1/2}$, $k(Q^2)$ will open upwards and $Q_{\text{sa}}^2>Q_{\text{max}}^2$, which means $k(Q^2)$ is positive on the interval $(0,Q_{\text{max}}^2)$.
	\item If $r_+/r_B=e^{-1/2}$, $k(Q^2)$ will be a linear function with respect to $Q^2$ and thus $k(Q^2)$ is always positive on the interval $(0,Q_{\text{max}}^2)$.
	\item If $r_+/r_B<e^{-1/2}$, $k(Q^2)$ will open downwards and $Q_{\text{sa}}^2<Q_{\text{max}}^2$, which means $k(Q^2)$ is positive on the interval $(0,Q_{\text{max}}^2)$.
\end{itemize}
Therefore, we conclude $C_{r_B,Q}\ge0$.

\subsection{$C_{r_B,\Phi}\ge0$}

To obtain the heat capacity at constant $r_B$ and $\Phi$, we first use Eq. (\ref{phiq}) to give $Q(r_+,r_B,\Phi,l)$, then we insert it into the expression of temperature (\ref{temq}). A straightforward calculation yields
\begin{equation}
	\begin{aligned}
		&\frac{\partial T}{\partial r_+}=A\left(a\Phi^4+b\Phi^2+c\right),\\
		&A=\frac{\left(\left(1-x^2\right)\left(\Phi^2-2\ln{x}\right)\right)^{-3/2}\left(-\ln{x}\right)^{-5/2}}{8\sqrt{2}\pi lx^2}\textcolor{blue}{,}\\
		&a=2\ln{x}\left(x^4-1\right)+4x^2\ln{x}^2-3(x^2-1)^2,\\
		&b=8\ln{x}(x^2-1)^2+4\ln{x}^2(-x^4+1)-16x^2\ln{x}^3,\\
		&c=16x^2\ln{x}^4,\\
		&x=\frac{r_+}{r_B}\in(0,1),
	\end{aligned}
\end{equation}
where $A,a,b,c$ are always positive. It follows that the sign of $\partial T/\partial r_+$ is the same as the expression in the brackets of the first line, which can be regarded as a quadratic function of $\Phi^2$. The symmetry axis of the quadratic function $-b/2a<0$, which implies that, for any given $\Phi$ the quadratic function is always positive since it opens upwards and $\Phi^2$ is non-negative. Therefore, we have $C_{r_B,\Phi}\ge0$.

In a canonical ensemble, a stable equilibrium states that the internal energy is a locally minimum against the virtual variation $\delta S$, and hence the positive heat capacities are sufficient to ensure that the non-extreme black holes are locally thermodynamically stable. In a grand canonical ensemble, a stable equilibrium states that the Gibbs free energy is a locally minimum against the virtual variation $(\delta S,\delta V)$. A sufficient condition for this is the second-order differential $\delta^2 G>0$. This could be transformed into requiring $\delta^2 E>0$. Equivalently, the Hessian matrix of $\delta^2 E$ in coordinates $(\delta S, \delta V)$ should be positive definite.

For the rotating BTZ black hole in a cavity, the Hessian matrix of $\delta^2 E$ in coordinates $(S,J)$ is
\begin{equation}
	H_{ij}^{R}=\left(
        \begin{array}{cc}
			\left(\frac{\partial T}{\partial S}\right)_J & \left(\frac{\partial T}{\partial J}\right)_S \\
			\left(\frac{\partial \omega}{\partial S}\right)_J & \left(\frac{\partial \omega}{\partial J}\right)_S \\
		\end{array}
		\right).
\end{equation}
The first diagonal element $(\frac{\partial T}{\partial S})_J=\frac{1}{4\pi}(\frac{\partial T}{\partial r_+})_J$, which has been proved to be positive. The second diagonal element
\begin{equation}
	\left(\frac{\partial \omega}{\partial J}\right)_S=\frac{4\left(r_B^2-r_+^2\right)}{\left(-J^2 l^2+4r_+^2 r_B^2\right)\sqrt{J^2\left(\frac{1}{r_B^2}-\frac{1}{r_+^2}\right)+\frac{4\left(r_B^2-r_+^2\right)}{l^2}}},
\end{equation}
which can be easily verified to be positive by using the condition $r_+<r_B$ and $J\le 2r_+^2/l$. The determinant of the Hessian matrix
\begin{equation}
	\left(\frac{\partial T}{\partial S}\right)_J \left(\frac{\partial \omega}{\partial J}\right)_S-\left(\frac{\partial T}{\partial J}\right)_S \left(\frac{\partial \omega}{\partial S}\right)_J=\frac{r_B^2\left(J^2 l^2-4r_+^4\right)}{16\pi^2 r_+^4\left(-r_B^2+r_+^2\right)\left(-J^2 l^2+4r_+^2 r_B^2\right)}.
\end{equation}
Notice that determinant is only positive when $J<2r_+^2/l^2$, that is, $T\neq0$. Therefore, we conclude that the non-extreme rotating BTZ black hole in a cavity is locally thermodynamically stable.

\begin{figure}[ptb]
	\begin{center}
		\includegraphics[width=0.48\textwidth]{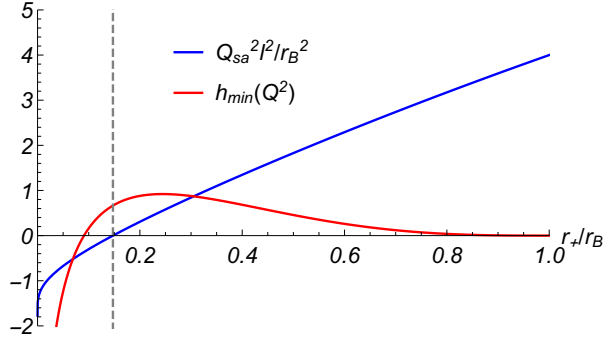}
		\caption{Plots of $Q_{\text{sa}}^2 l^2/r_B^2$ and $h_{\text{min}}(Q^2)$ with respect to $r_+/r_B$.}
		\label{comp}
	\end{center}
\end{figure}

For the charged BTZ black hole in a cavity, the Hessian of the second order differential of energy $\delta^2 E$ in coordinates $(S,Q)$ is
\begin{equation}
	H_{ij}^{C}=\left(
	\begin{array}{cc}
		\left(\frac{\partial T}{\partial S}\right)_Q & \left(\frac{\partial T}{\partial Q}\right)_S \\
		\left(\frac{\partial \Phi}{\partial S}\right)_Q & \left(\frac{\partial \Phi}{\partial Q}\right)_S \\
	\end{array}
	\right).
\end{equation}
The first diagonal element is positive, and the second diagonal element reads
\begin{equation}
	\left(\frac{\partial \Phi}{\partial Q}\right)_S=\frac{2\sqrt{2}\left(-r_B^2+r_+^2\right)\ln{\frac{r_+}{r_B}}}{l^2\left(\frac{2\left(r_B^2-r_+^2\right)}{l^2}+Q^2\ln{\frac{r_+}{r_B}}\right)^{3/2}},
\end{equation}
which is also positive. The determinant of the Hessian is
\begin{equation}
	\left(\frac{\partial T}{\partial S}\right)_Q \left(\frac{\partial \Phi}{\partial Q}\right)_S-\left(\frac{\partial T}{\partial Q}\right)_S \left(\frac{\partial \Phi}{\partial S}\right)_Q=\frac{8l^2 Q^2\left(-r_B^2+r_+^2\right)-\left(l^2 Q^2+4r_B^2\right)\left(l^2 Q^2+4r_+^2\right)\ln{\frac{r_+}{r_B}}}{32\pi^2 r_+^2\left(2\left(r_B^2-r_+^2\right)+l^2 Q^2\ln{\frac{r_+}{r_B}}\right)^2}.
\end{equation}
The denominator of the expression is apparently positive. The numerator can be regarded as a quadratic function of $Q^2$ and we denote it as $h(Q^2)$. The symmetric axis of $h(Q^2)$ is given by
\begin{equation}
	Q_{\text{sa}}^2=-\frac{2\left(r_B^2+r_+^2\right)}{l^2}-\frac{4\left(r_B^2-r_+^2\right)}{l^2 \ln{\frac{r_+}{r_B}}}.
\end{equation}
The quadratic function $h(Q^2)$ opens upward, so the minimal value of $h(Q^2)$ is
\begin{equation}
	h_{\text{min}}(Q^2=Q_{\text{sa}}^2)=16\left(r_B^4-r_+^4\right)+\frac{4\left(r_B^2-r_+^2\right)^2\left(4+\left(\ln{\frac{r_+}{r_B}}\right)^2\right)}{\ln{\frac{r_+}{r_B}}}.
\end{equation}	
We plot the curves of $Q_{\text{sa}}^2 l^2/r_B^2$ and $h_{\text{min}}(Q^2)$ with respect to $r_+/r_B$ in FIG. \ref{comp}, where we find that the values of $r_+/r_B$ that make $Q_{\text{sa}}^2$ positive also make $h_{\text{min}}(Q^2)$ positive. That is to say, $h_{\text{min}}(Q^2)$ has to be positive since $Q_{\text{sa}}^2$ must be positive. Therefore, we conclude that the system of a charged BTZ black hole in a cavity is locally thermodynamically stable.

\section{Phase Transitions of BTZ Black Hole without Cavity}\label{ptbbgc}

It is of great interest to investigate the phase transitions of a BTZ black hole without a cavity so as to make a comparison with the case of the cavity existing. For a asymptotically flat black hole in a cavity, thermodynamic quantities without a cavity could be straightforwardly obtained by imposing $r_B\to\infty$. However, this does not hold for the BTZ black hole whose spacetime is not asymptotically flat. For the static and neutral one, the thermodynamic quantities are
\begin{equation}
	\begin{aligned}
		T&=\frac{1}{4\pi}\left(\frac{2r_+}{l^2}\right),\\
		F&=-E=-\frac{r_+^2}{l^2}.
	\end{aligned}
\end{equation}
The free energy of thermal AdS space is
\begin{equation}
	F_{AdS}=-1,
\end{equation}
Thus the critical temperature is given by
\begin{equation}
	T_c=\frac{1}{2\pi l}.
\end{equation}

\begin{figure}[ptb]
	\begin{center}
		\subfigure{
			\includegraphics[width=0.35\textwidth]{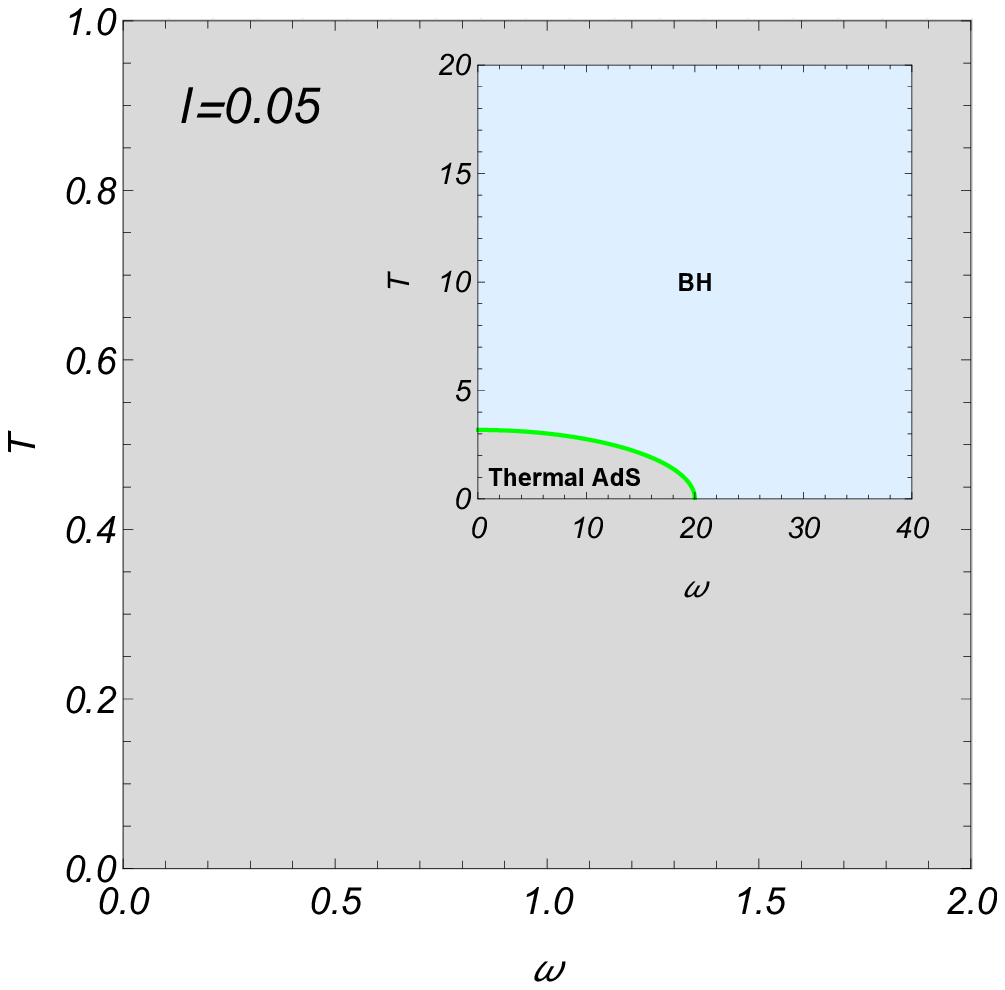}}
		\subfigure{
			\includegraphics[width=0.35\textwidth]{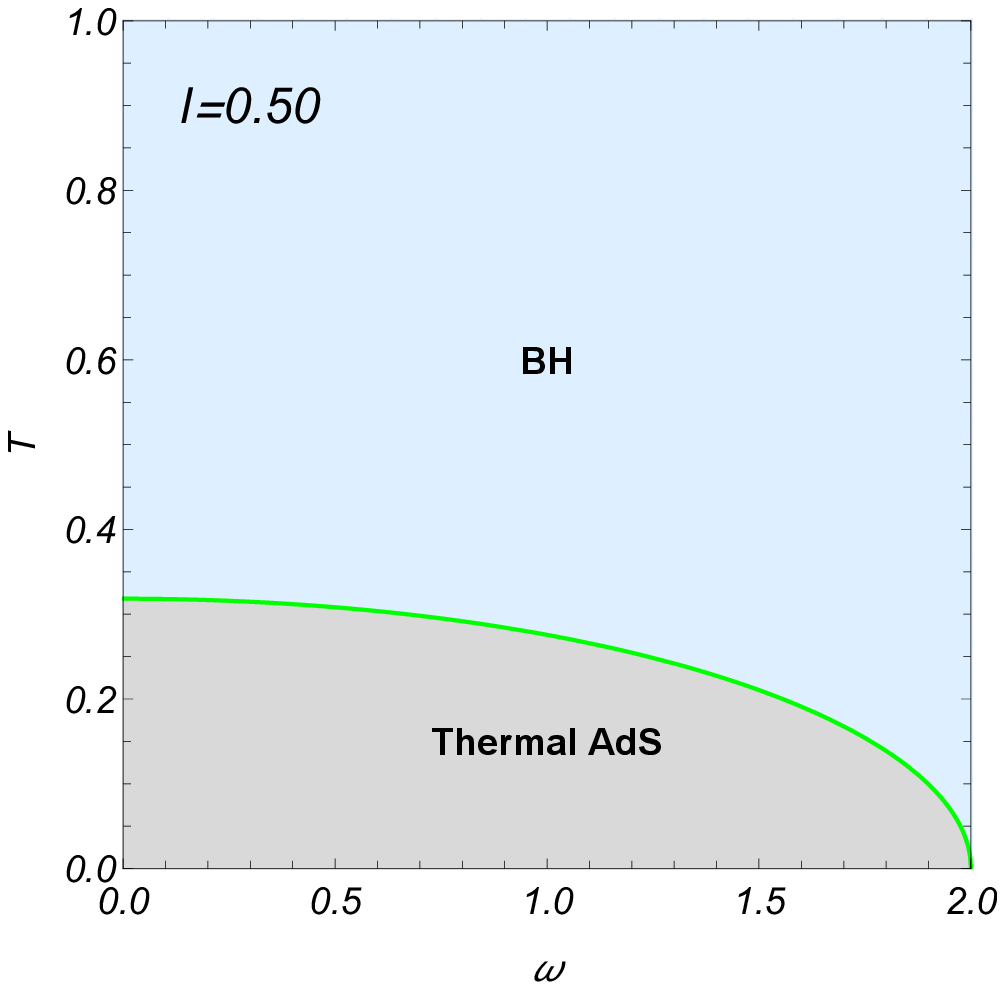}}
		
		\subfigure{
			\includegraphics[width=0.35\textwidth]{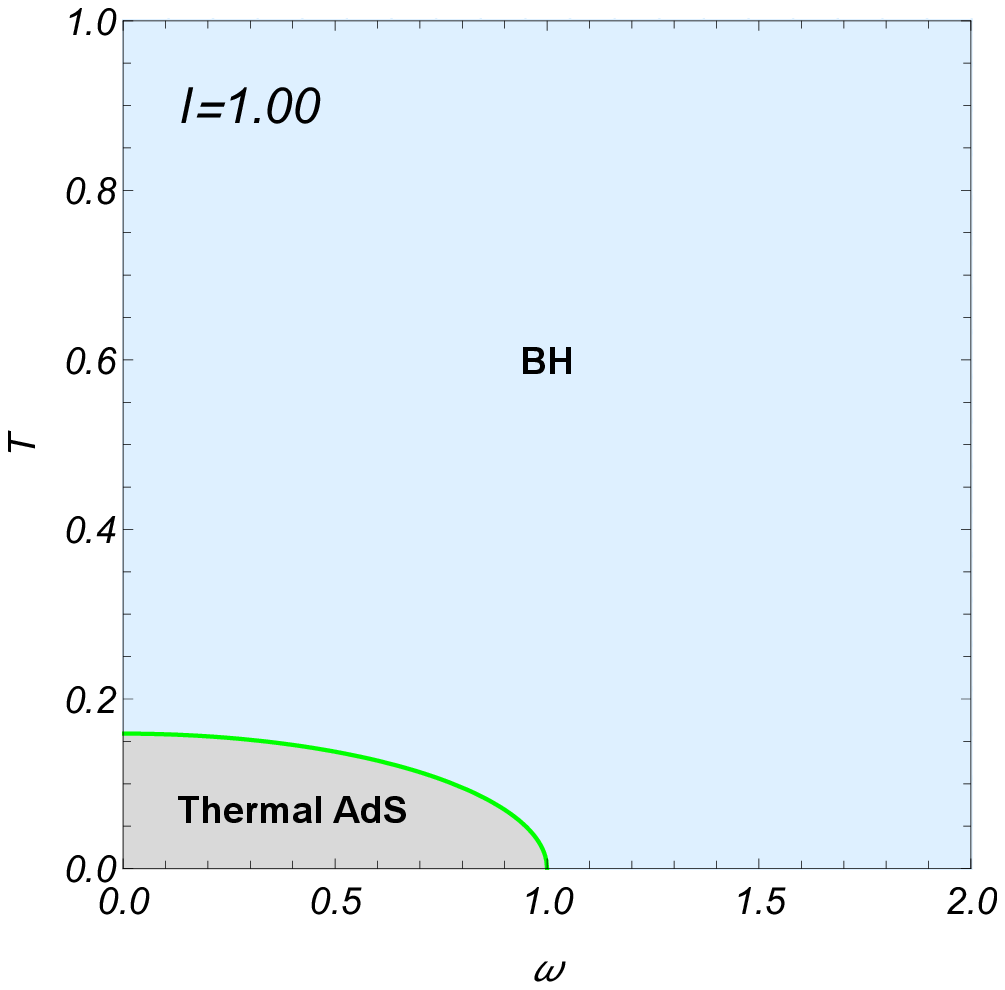}}
		
		\subfigure{
			\includegraphics[width=0.35\textwidth]{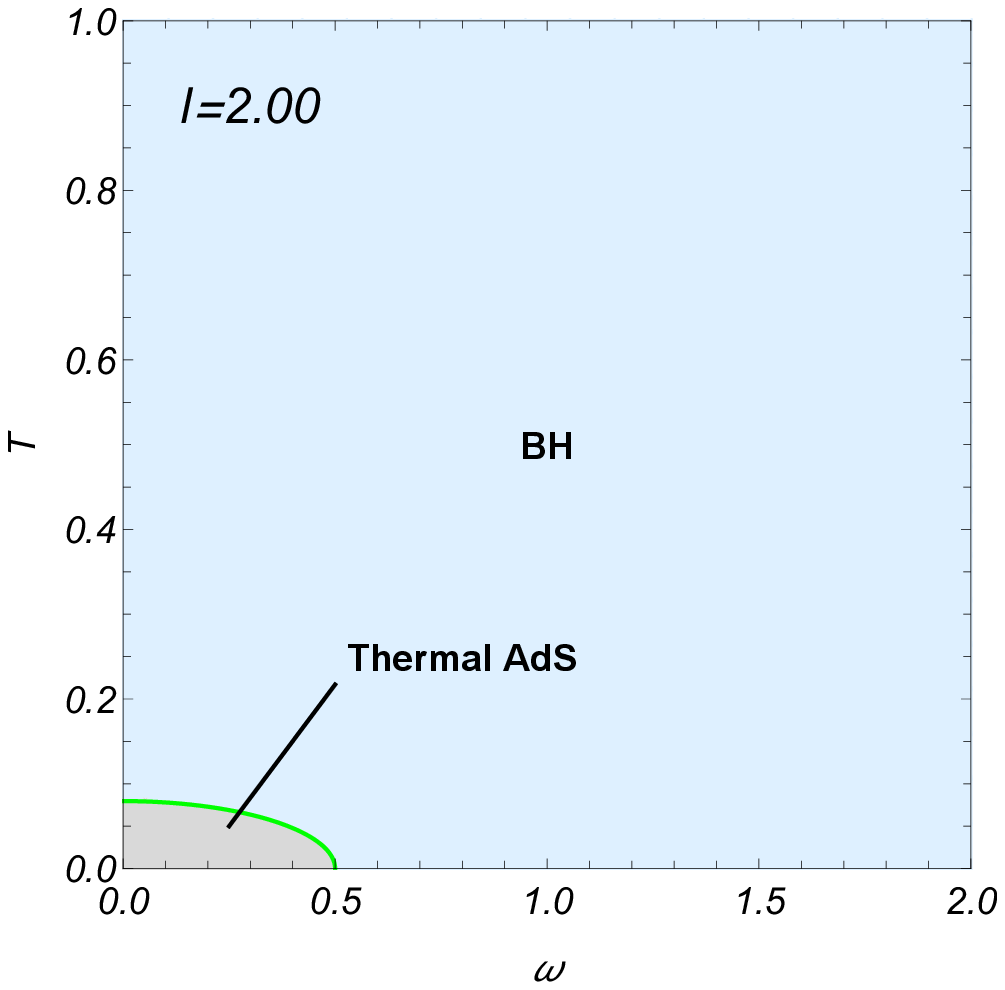}}
		\subfigure{
			\includegraphics[width=0.35\textwidth]{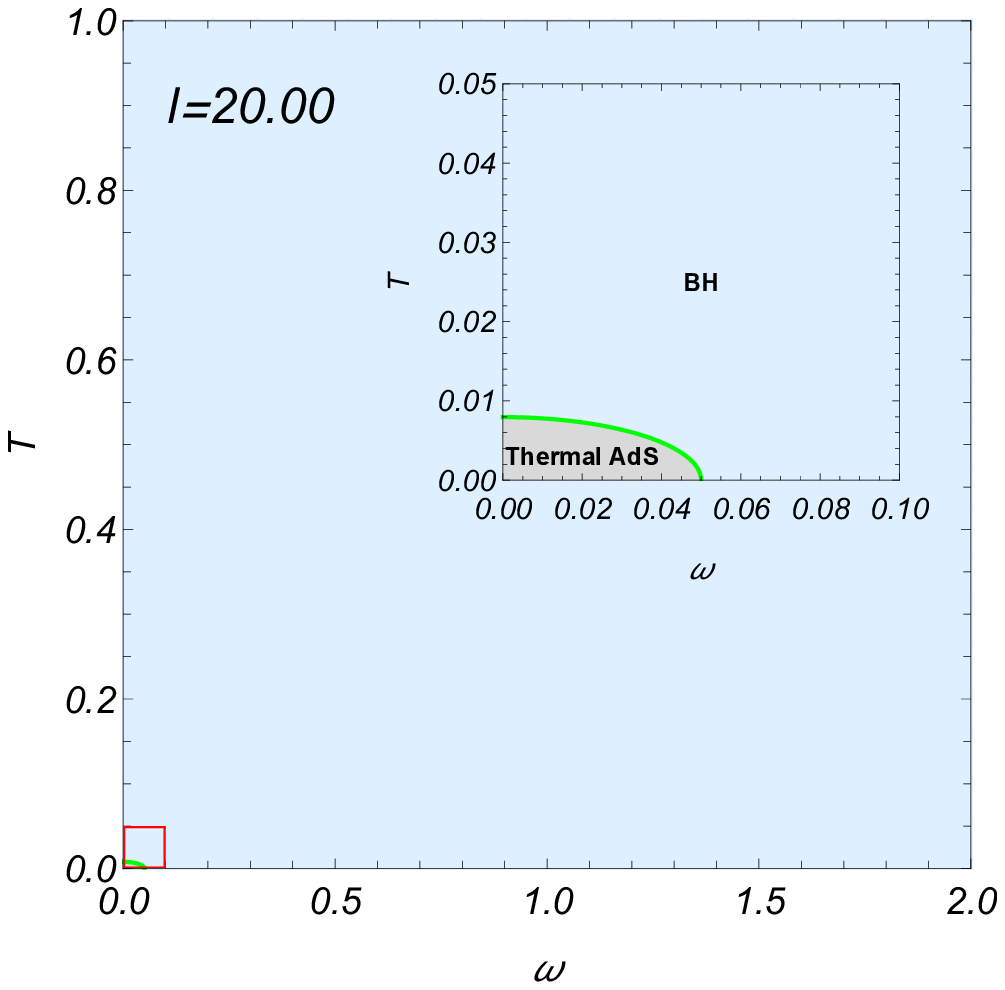}}
		\caption{Phase diagrams of the rotating BTZ black hole without the cavity. The AdS radius $l=0.05,\text{ }0.50,\text{ }1.00,\text{ }2.00,\text{ }20.00$, respectively. The green curve is a first-order phase transition curve.}
		\label{NocavityJ}
	\end{center}
\end{figure}

Taking the angular momentum into consideration, thermodynamic quantities can be written as
\begin{equation}
	\begin{aligned}
	T&=\frac{1}{4\pi}\left(\frac{2r_+}{l^2}-\frac{J^2}{2r_+^3}\right),\\
	G&=-\frac{r_+^2}{l^2}+\frac{J^2}{4r_+^2},\\
	E&=\frac{r_+^2}{l^2}+\frac{J^2}{4r_+^2}\textcolor{blue}{,}\\
	\omega&=\frac{J}{2r_+^2}.
	\end{aligned}
\end{equation}
Notice that the additional energy brought by angular momentum can be linearly added into the total energy which is different from the case in a cavity. The phase diagrams of a rotating black hole are exhibited in FIG. \ref{NocavityJ}.

\begin{figure}[ptb]
	\begin{center}
		\subfigure{
			\includegraphics[width=0.35\textwidth]{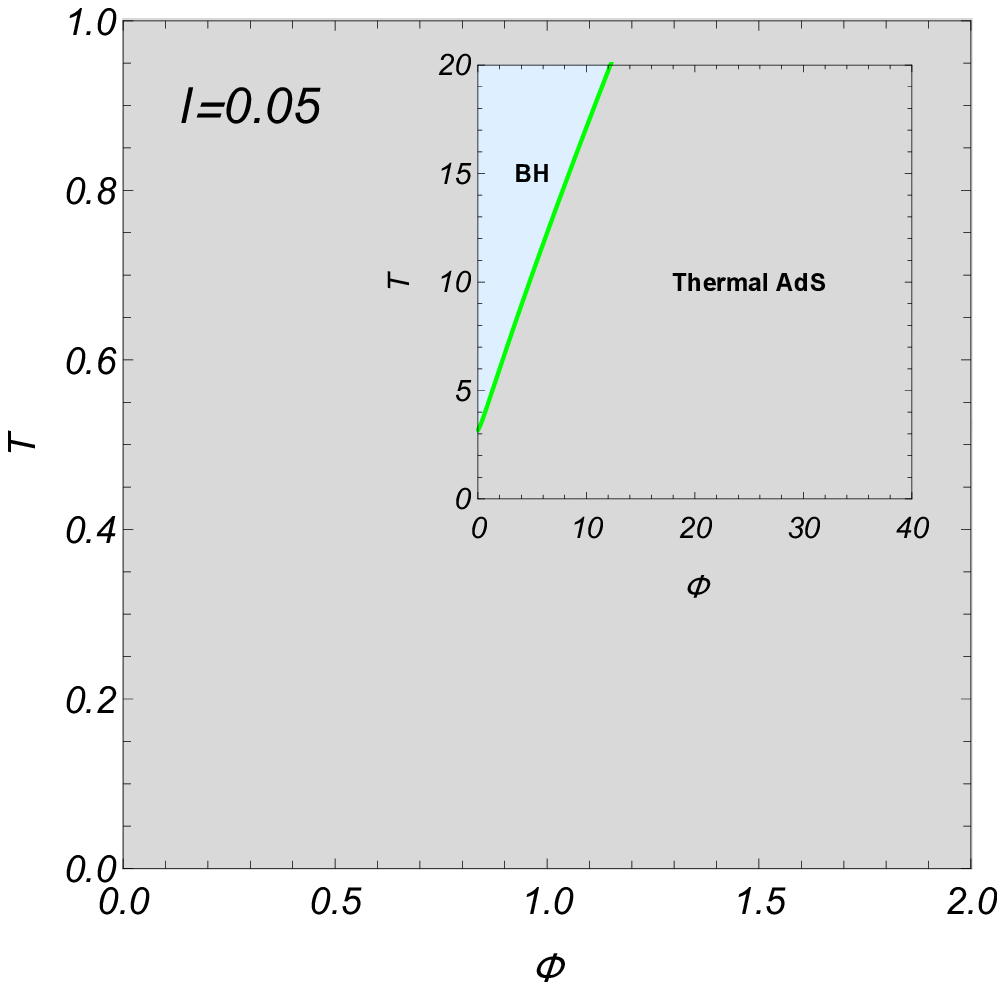}}
		\subfigure{
			\includegraphics[width=0.35\textwidth]{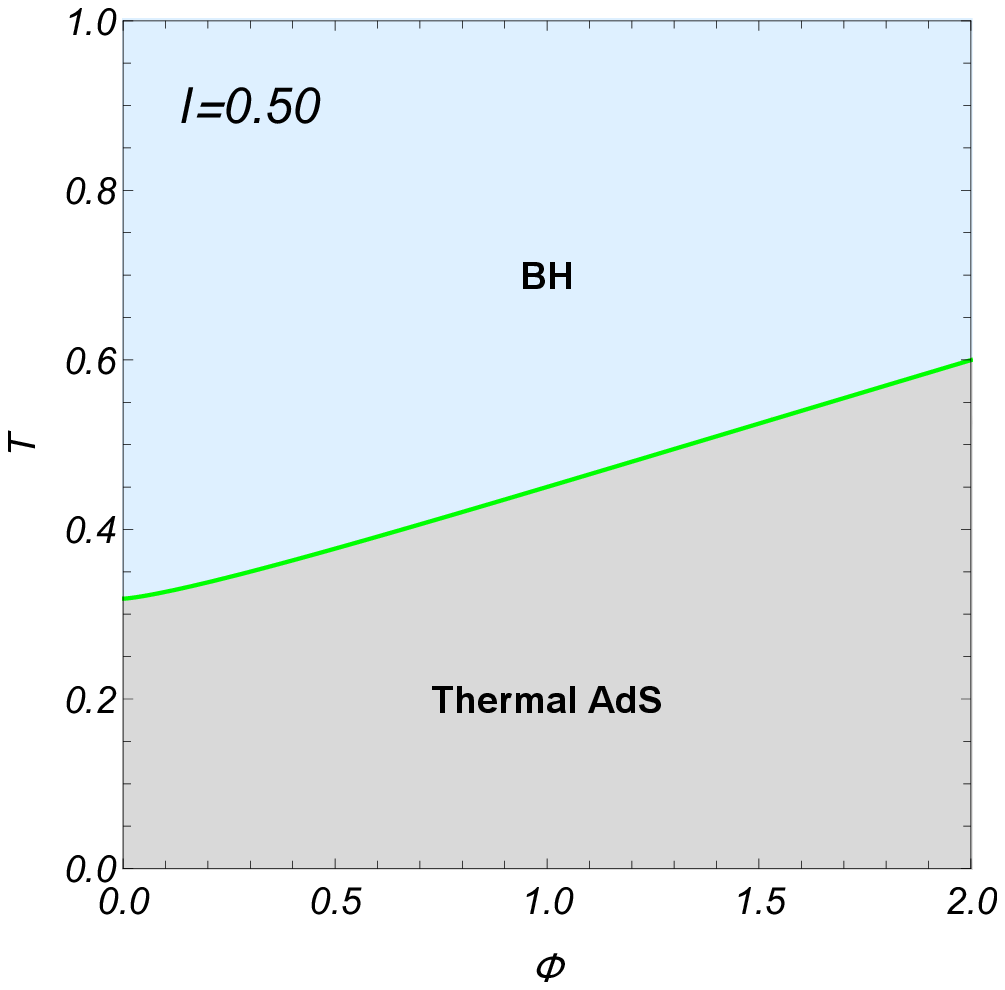}}
		
		\subfigure{
			\includegraphics[width=0.35\textwidth]{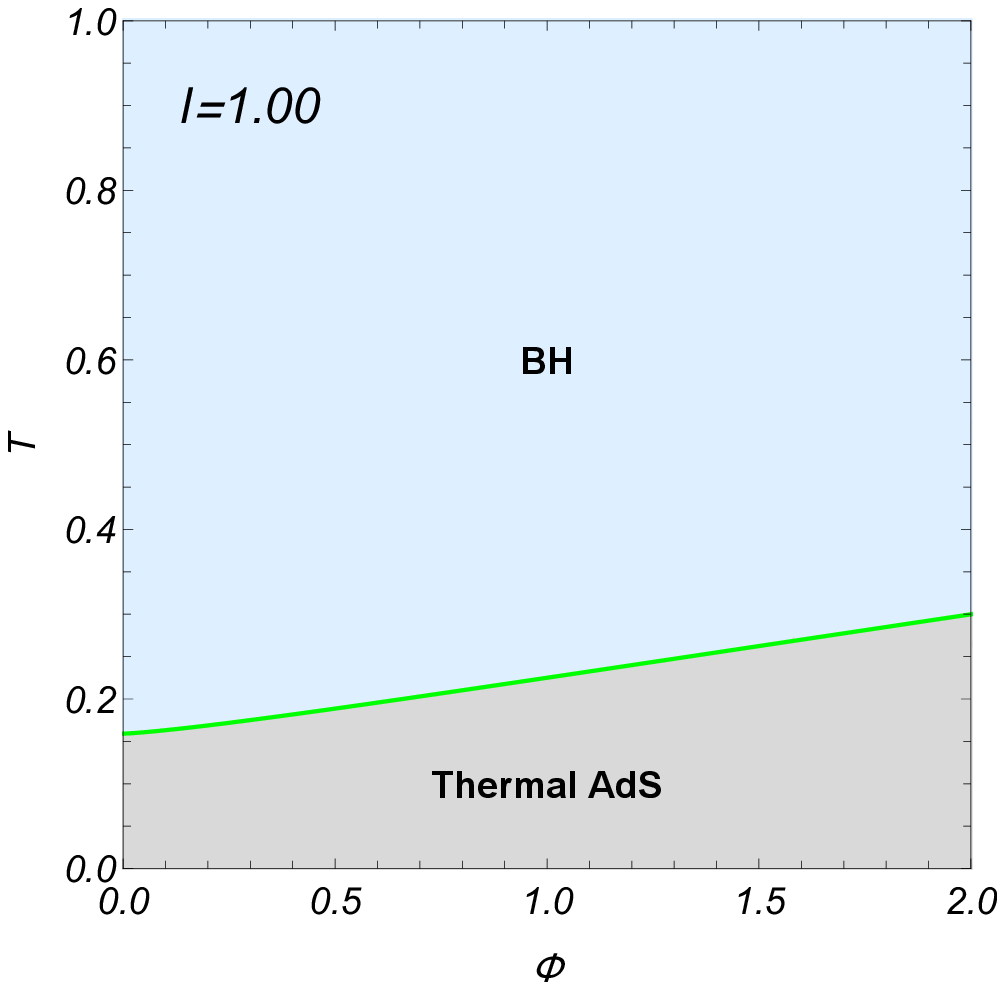}}
		
		\subfigure{
			\includegraphics[width=0.35\textwidth]{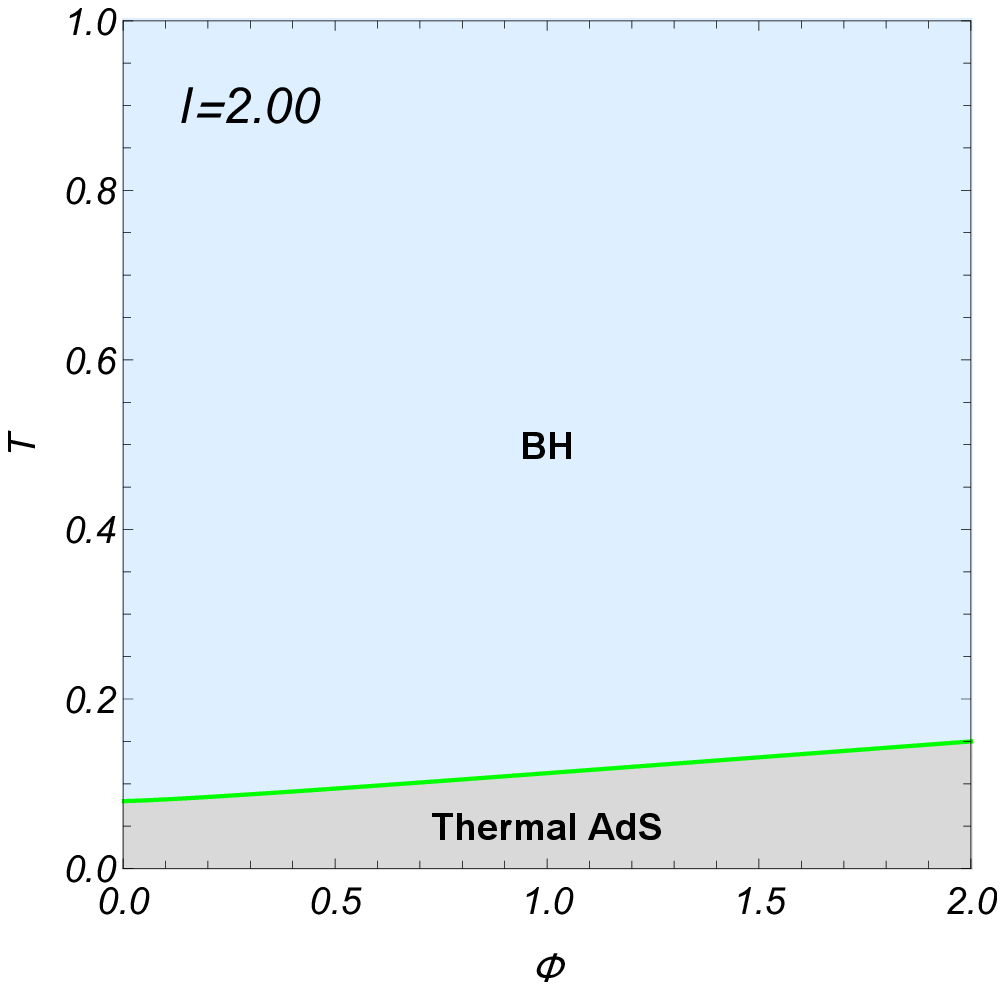}}
		\subfigure{
			\includegraphics[width=0.35\textwidth]{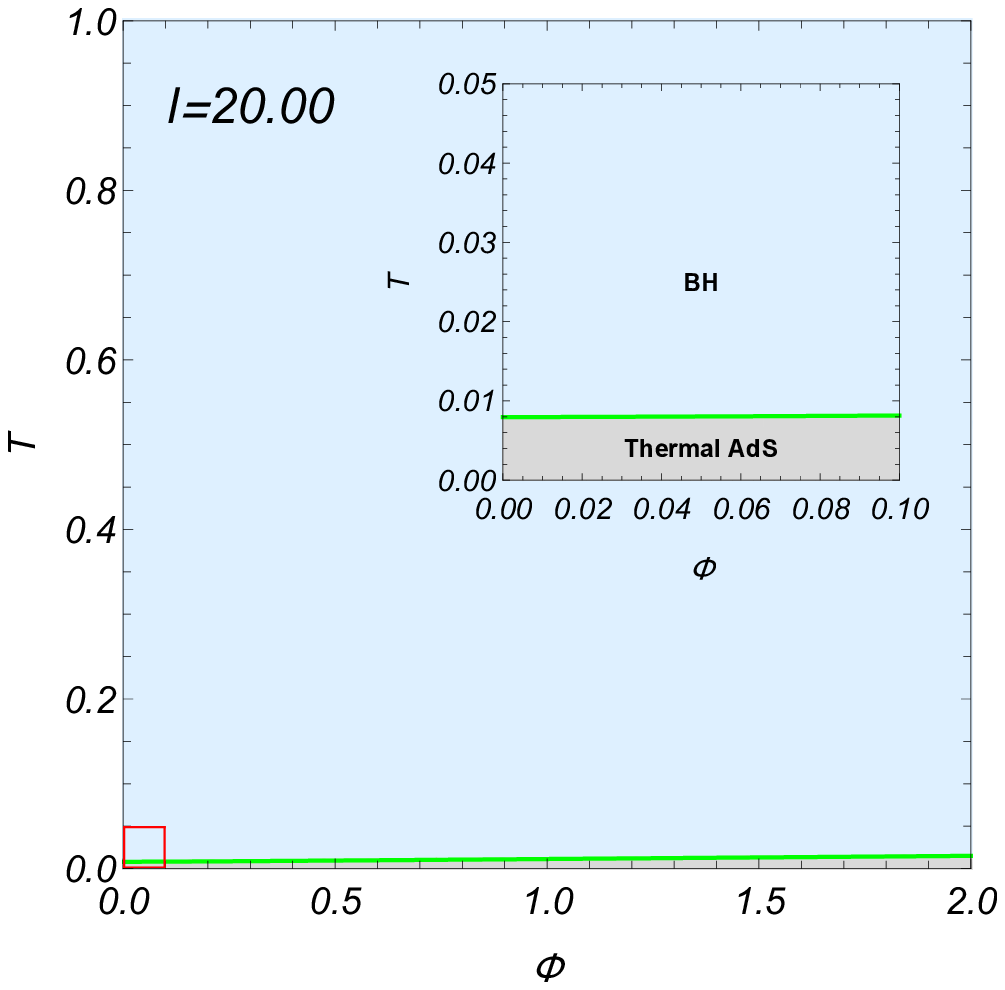}}
		\caption{Phase diagrams of the charged BTZ black hole without the cavity. The AdS radius $l=0.05,\text{ }0.50,\text{ }1.00,\text{ }2.00,\text{ }20.00$, respectively. The green curve is a first-order phase transition curve.}
		\label{NocavityQ}
	\end{center}
\end{figure}

The thermodynamic quantities of a charged BTZ black hole are
\begin{equation}
	\begin{aligned}
		T&=\frac{1}{4\pi}\left(\frac{2 r_+}{l^2}-\frac{Q^2}{2 r_+}\right),\\
		G&=-\frac{r_+^2}{l^2}+\frac{Q^2}{2}\ln{\frac{r_+}{l}}+\frac{Q^2}{2},\\
		E&=\frac{r_+^2}{l^2}-\frac{Q^2}{2}\ln{\frac{r_+}{l}}\textcolor{blue}{,}\\
		\Phi&=-Q\ln{\frac{r_+}{l}}.
	\end{aligned}
\end{equation}
The phase diagrams are shown in FIG. \ref{NocavityQ}.

\normalem
\bibliographystyle{unsrturl}
\bibliography{Cavity}

\end{document}